\documentclass[pagesize,12pt,a4paper]{scrartcl}
\usepackage{inputenc}
\usepackage[T1]{fontenc}
\usepackage{mathptmx}
\usepackage{helvet}
\usepackage{courier}
\usepackage{type1cm}   
\usepackage[bottom]{footmisc} 
\usepackage[pstricks,squaren,cdot,thickqspace]{SIunits}
\usepackage[english]{babel}
\usepackage{mathtools}
\usepackage{mathdots}
\usepackage{amsmath}
\usepackage{amssymb}
\usepackage{amsxtra}
\usepackage{amsfonts}
\usepackage[squaren]{SIunits}
\usepackage{graphicx}
\usepackage{multirow}
\usepackage{mathrsfs}
\usepackage{rotating}
\usepackage{appendix}
\usepackage{textcomp}
\usepackage{manyfoot}
\usepackage{booktabs}
\usepackage{capt-of}
\usepackage{xcolor}
\usepackage{braket}
\usepackage{url}
\usepackage{geometry}
\usepackage{wrapfig}
\usepackage{comment}

\newcommand{\ir}{\mathrm{i}}
\newcommand{\e}{\mathrm{e}}

\newcommand{\eins}{{\mathbf 1}}

\renewcommand{\jmath}{j}
\newcommand{\longpage}{\enlargethispage{1\baselineskip}}
\newcommand*\idx[1]{\index{#1}#1}

\providecommand{\abs}[1]{\lvert#1\rvert}

\newcommand{\clearemptypage}{\newpage{\pagestyle{empty}\clearpage}}

\DeclareMathOperator{\dist}{dist\,}

\DeclareMathOperator{\ch}{ch}
\DeclareMathOperator{\sh}{sh}

\DeclareMathOperator{\spec}{spec}

\DeclareMathOperator{\Ad}{Ad}

\DeclareMathOperator{\dv}{d}

\DeclareMathOperator{\T}{T}

\DeclareMathOperator{\dt}{\tilde d\!}

\newcommand{\osumS}{\parbox {0pt} {\kern0.1ex\makebox[0pt][l]{$\,\textstyle \circ $}}\sum}
\newcommand{\osumSS}{\parbox {0pt} {\kern0.15ex\makebox[0pt][l]{$\,\textstyle \circ$}}\sum}

\ifx\KOMAScript\undefined%
  \DeclareRobustCommand{\KOMAScript}{\textsf{K\kern.05em O\kern.05em%
      M\kern.05em A\kern.1em-\kern.1em Script}}
\fi

\newlength{\help}
\newlength{\minuslaenge}
\begin{document}

\title{Bundle Structure of Massless Unitary Representations of the Poincaré Group}
\author{Norbert Dragon \\
Institut für Theoretische Physik\\ Leibniz Universität Hannover\\
orcid 0000-0002-3809-524X
}
\date{}

\maketitle




\abstract{Reviewing the construction of induced representations of the Poincaré group of four-dimensional spacetime
we find all massive representations, including the ones acting on interacting many-particle states. Massless momentum wavefunctions
of non-vanishing helicity turn out to be more precisely sections of a U$(1)$-bundle over the massless shell, a property which to date was overlooked
in bracket notation. Our traditional notation enables questions about square integrability and smoothness. Their answers 
complete the picture of relativistic quantum physics.

Frobenius' reciprocity theorem  prohibits massless one-particle states with total angular momentum less than the modulus of the helicity.
There is no two-photon state with $J=1$, explaining the longevity of orthopositronium.

Partial derivatives of the momentum wave functions are no operators which can be applied to massless states $\Psi$
with nonvanishing helicity. They allow only for covariant, noncommuting derivatives. The massless shell has a noncommutative
geometry with helicity  being its topological charge. A spatial position operator for $\Psi$, 
which constitutes Heisenberg pairs with the spatial momentum, is excluded by the smoothness requirement of the domain of the Lorentz generators. 

}

\clearemptypage

\section{Introduction}

Some presentations of relativistic quantum physics, in particular text books on string theory, postulate the generators 
of the Poincaré group in Hilbert space to be given in terms of position operators 
$X=(X^0,X^1,\dots ,X^{D-1})$ and
momentum operators $P=(P^0,P^1,\dots , P^{D-1})$, which generate a $2D+1$-dimensional Heisenberg 
group of operators $U_a V_b \e^{\ir c}$, ($a,b\in \mathbb R^D$, $c\in \mathbb C$),
\begin{equation}
\label{weyl}
U_a = \e^{\ir P a}\ ,\ V_b= \e^{\ir X b}\ ,\ U_aU_b = U_{a+b}\ ,\ V_aV_b = V_{a+b}\ ,\ 
U_a V_b = V_b U_a \e^{\ir a \cdot b},
\end{equation}
where $a\cdot b = \eta^{mn}a_m b_n = a_0 b_0 - a_1 b_1 - \dots -a_{D-1} b_{D-1}$. 
On smooth and rapidly decreasing momentum wave functions $\Psi: \mathbb R^D \rightarrow \mathcal V\,,\,p\mapsto \Psi(p)$,
where $\mathcal V$ is a sum of finite dimensional spaces with
matrix representations of the Lorentz group and a non-degenerate scalar product,
the operators $P$ and $X$ act by multiplication and differentiation
\begin{equation}
(P^m\Psi)(p)= p^m \Psi(p)\ ,\ (X^n \Psi)(p) = - \ir \eta^{nm}\partial_{p^m}\Psi(p)\ ,
\end{equation}
hermitian with respect to the scalar product
\begin{equation}
\braket{\Phi | \Psi}= \int\!\dv^D\!\!p\  \bigl(\Phi(p)|\Psi(p)\bigr)_{\mathcal V}\ ,
\end{equation}
and represent the Heisenberg algebra, which in covariant string theory is part of its canonical quantization,
\begin{equation}
\label{heisen}
 [P^m, P^n]= 0 = [X^m, X^n]\ ,\ [P^m, X^n] = \ir \eta^{mn}\eins\ .
\end{equation}
Hence, given finite dimensional matrices $\Gamma_{mn}$ which commute with $X$ and~$P$ and generate re\-pre\-sen\-tations
of the Lorentz group in $\mathcal V$,
the operators 
\begin{equation}
\label{lorstring}
-\ir M_{mn} \stackrel{?}{=}-\ir\bigl(X_m P_n - X_n P_m\bigr) + \Gamma_{mn}
\end{equation}
represent, analogous to angular momentum operators, the Lorentz algebra\footnote{Notation: 
Let $T_{a}: x \mapsto x + a$ denote a translation in $\mathbb R^4$, $T_\Lambda: x \mapsto \Lambda x$ a Lorentz transformation, 
and $T_{a,\Lambda}=T_a T_\Lambda \in \mathfrak P$ a Poincar\'e transformation.
We denote by $U_{a,\Lambda}$ its unitary representation with generators $P^m$ and $M_{mn}= -M_{nm}$, 
$U_{a,\e^\omega}= \e^{\ir P a}\e^{\ir \omega^{mn }M_{mn}/2}$,
in a Hilbert space of one-particle states. }
\begin{equation}
\label{loralgebra}
 [\Gamma_{kl},\Gamma_{mn}]= - \eta_{km}\Gamma_{ln}+ \eta_{kn}\Gamma_{lm} + \eta_{lm}\Gamma_{kn}- \eta_{ln}\Gamma_{km}\ .
\end{equation}

\longpage

Prevalent as this representation is, 
it is \emph{fundamentally flawed:} 
the Weyl relations (\ref{weyl}) of $X^0$ require $P^0$ 
to be unitarily equivalent to $P^0 + b$, $b\in \mathbb R$, 
\begin{equation}
\e^{\ir b X^0}\,P^0\, \e^{-\ir b X^0}= P^0 + b\ .
\end{equation}
So the spectrum of $P^0$ is translation invariant, $\spec(P^0)=\spec(P^0 + b)$.

In all acceptable relativistic theories $\spec(P^0)$ is non-negative. Hence
they allow neither $X^0$ nor any operator $M_{0n}$ (\ref{lorstring}) which employs~$X^0$.
Postulating constraints to restrict the spectrum to positive energy mass shells leads to contradictions \cite{dragono1}. 
The Heisenberg algebra is incompatible with a unitary representation of the Poincar\'e group $\mathfrak P$ on one-particle states.

Another reason to take the Heisenberg algebra as basis for the construction of Lorentz generators may be their
action on fields~$\Phi(x)$ with
\begin{equation}
U_{a,\Lambda} \Phi(x) U_{a,\Lambda}^{-1} = \Phi(\Lambda^{-1}(x-a))\ .
\end{equation}
But neither $x$ nor $ \partial_x$ nor $\Phi(x)$ are operators in a Hilbert space. The fields are operator valued distributions.
Integrated with smooth, rapidly decreasing testfunctions~$f$
\begin{equation}
\Phi_f = \int\!\! \dv^4\! x\, f(x)\,\Phi(x)
\end{equation}
and applied to the vacuum they generate a set of one-particle states $\Phi_f \Omega$ which, by the Reeh-Schlieder theorem \cite{reeh}, 
is dense already if all $f$ are restricted  to vanish outside some arbitrarily  chosen fixed open set. So the argument~$x$ of the fields
is not precisely the position of a particle, however suggestive this language about Feynman graphs may be.

To remind the reader we review the construction of induced representations. In appendices we provide the angle of 
the Wigner rotations in the massive and massless cases, filling the gap which prevented to date the straight forward 
determination of the Lorentz generators. 

In the massive case they act on a sum of Hilbert spaces of 
fixed spin $s$, $(2s+1)\in \mathbb N$, which all have the peculiarity to factorize
\begin{equation}
\label{hred}
\mathcal H = \sum_s \mathcal H_s\ ,\ 
\mathcal H_s = \hat{\mathcal H_{s}} \otimes \mathcal I_s
\end{equation}
into  a Hilbert space $\hat{\mathcal H_{s}}$ which carries an irreducible spin-$s$ representation of $\mathfrak P$ with \emph{unit mass} times
a Hilbert space $\mathcal I_s$ which is pointwise invariant and is acted upon by the positive mass operator $M=\sqrt{P^2}$.

One-particle representations act on the span of eigenspaces of $M$ and are irreducible if and only if~$\mathcal I_s = \mathbb C$. 

This settles the question, what an interacting representation should be. For it necessary conditions have been specified \cite{weinberg}
but no solution. For fixed spin massive representations of $\mathfrak P$ can differ from the free representation with mass $M$ only by 
employing an interacting mass $M'$. On many-particle states it acts in~$\mathcal I_s$ as Hamiltonian of the relative motion \cite{dragon3}.

Massless representations are \emph{not} a continuous limit $m \rightarrow 0$ of massive representations. Their mass shell encloses a Lorentz fixed point $p=0$
and has the topology $\mathbb R \times S^{D-2}$. Therefore the generators do not act on smooth wavefunctions of $\mathbb R^{D-1}$ but, as it turns out, on sections of
bundles with nontrivial transition functions. We derive these results by elementary analysis from the requirement of square integrability of rotated states.

The helicity of a photon cannot combine with its orbital angular momentum to a rotation invariant state: \lq There is no round photon\rq. 
Consequently there is no two-photon state which combines the helicities and the relative orbital angular momentum to $j=1$, explaining why positronium with
$j=1$ cannot decay into two photons \cite{landau, yang}.

We also show that no spatial position operator $\vec X$ can exist which 
constitutes Heisenberg pairs with the spatial momenta and generates translations of spatial momentum.

\section{Induced Representations}

Mackey's theorems \cite{mackey} classify the unitary, strongly\index{strong limit}
measurable\footnote{For each $\Psi \in \mathcal H$ the map $f_\psi: \mathfrak P \rightarrow \mathcal H,\, g \mapsto U_g \Psi$ 
has to be measurable.}  
representations $U\!\! : g\mapsto U_g$ of the Poincaré group~$\mathfrak P$ of a $D$-dimensional spacetime
in a Hilbert space $\mathcal H$, $U_g: \mathcal H \rightarrow \mathcal H$, $U_g U_{g'}=U_{g\,g'}$.

Each $U$ decomposes, uniquely up to unitary equivalence, 
into a sum of representations which act on a product of $\hat{\mathcal H_s}$ with an irreducible representation, and
a pointwise invariant space $\mathcal I_s$ (\ref{hred}). 

To describe the irreducible representations in the cases of interest let $G$ denote the cover of the restricted Lorentz group and $H\subset G$ 
the cover of the stabilizer of some fixed timelike or lightlike momentum $\underline p\in \mathbb R^D$ 
\begin{equation}
H=\Set{ h\in G: h \underline p = \underline p}\ .
\end{equation}
The Lorentz orbit of $\underline p$ is a mass shell $\mathcal M\sim G/H$ with $m > 0$ or $m=0$, 
\begin{equation}
\label{massshell}
\mathcal M_m = \Set{ 
\begin{pmatrix}
\sqrt{m^2+\vec p^2}\\
\vec p
\end{pmatrix}:\vec p \in \mathbb R^{D-1} }\ , \  
\mathcal M_0 = \Set{ 
\begin{pmatrix}
\sqrt{\vec p^2}\\
\vec p
\end{pmatrix}:\vec p \ne 0 
 }\ .
\end{equation}
Their points correspond one-to-one to the left cosets $gH$, as $p = g \underline p = gh \underline p $.
As these cosets are either identical or disjoint, $G$ is a bundle over a mass shell $\mathcal M$ with fibers, which are each diffeomorphic to $H$.

Let~$R$ represent $H$ unitarily in a Hilbert space $\mathcal V$, e.g. in $\mathbb C^{2s + 1}$,
\begin{equation}
R(h_2)\,R(h_1) = R(h_2h_1)\ ,
\end{equation}
and denote by $\mathcal H_R$ the space of functions $\Psi: G \rightarrow \mathcal V$ which satisfy 
\begin{equation}
\label{psifaser1}
\Psi(gh) = R(h^{-1})\,\Psi(g)\quad \forall g\in G,\, h\in H  \ .
\end{equation}
$\mathcal H_R$ is mapped to itself by the multiplicative representation of translations~$a$,
\begin{equation}
\label{mometumrep}
(U_a \Psi)(g) = \e^{\ir p a}\Psi(g)\ ,\ p = g \underline p\ ,
\end{equation}
and by composition with the inverse Lorentz transformations 
\begin{equation}
\label{psibuendel}
(U_g \Psi)(g^\prime) = \Psi(g^{-1}g^\prime)\ ,\ U_g \Psi = \Psi \circ g^{-1}\ .
\end{equation}
One easily confirms that $U_a$ and $U_g$ (\ref{mometumrep},\ref{psibuendel}) represent $\mathfrak P$, 
$U_a\,U_b = U_{a+b}$\,, $U_g\, U_{g^\prime}=U_{g\,g^\prime}$ and $U_g\, U_a = U_{b}\, U_g$ with $b = g a$.


As $R$ is unitary, the scalar product of the values at $g$ of two functions $\Phi, \Psi\in \mathcal H_R$, $\Phi(g), \Psi(g)\in \mathcal V$,  
\begin{equation}
\bigl(\Phi(g)|\Psi(g)\bigr)=\bigl(R(h^{-1})\,\Phi(g)|R(h^{-1})\,\Psi(g)\bigr)=
\bigl(\Phi(gh)|\Psi(gh)\bigr)
\end{equation}
is constant in each left coset $gH$. 
So one can define a scalar product in $\mathcal H_R$ 
\begin{equation}
\Braket{\Phi | \Psi} = \int_{\mathcal M} \dt p\, \bigl(\Phi| \Psi\bigr)_{p}
\end{equation}
where $\dt p = \dt\, (g\,p)$ is a Lorentz invariant measure. We choose it
with  the conventional normalization factors of quantum field theory $(m\ge 0)$,
\begin{equation}
\label{lorentzmaas}
\dt p = \frac{\dv^{\,3}\!\! p}{(2\pi)^3\, 2\,\sqrt{m^2+\vec p^2}}\ ,\ \dt\, (g p) = \dt p\ .
\end{equation}

Equipped with this scalar product $\mathcal H_R$ becomes a Hilbert space and
$U$ the unitary representation, 
\begin{equation}
\int_{\mathcal M}\!\! \dt\, (g p)\, \bigl(U_g \Phi| U_g \Psi\bigr)_{g p}
=\int_{\mathcal M}\!\! \dt\, (g p)\, \bigl( \Phi|  \Psi\bigr)_{p}
=\int_{\mathcal M}\!\! \dt p\, \bigl( \Phi|  \Psi\bigr)_{p}\ ,
\end{equation}
which one calls induced by the representation $R$ of the stabilizer $H$.

Two such induced representations are equivalent if and only if
the inducing representations $R$ are equi\-valent and their masses
coincide. They are irreducible only if $R$ is an irreducible representation of~$H$.

The states, on which an irreducible representation acts, are the possible states of
a single particle.  In sufficient distance to other particles its time evolution is generated 
for an observer at rest by $P^0$ and completely determined by its mass, no matter whether it is composite
or elementary  and whether it is charged or neutral.


\section{Sections of Bundles}

The momentum wave function of a spin-s particle in state $\psi$ has the physical meaning to give the probability 
\begin{equation}
w(\Delta, P ,\psi) =  \int_\Delta \dt p \, \sum_{n= -s}^s |\psi^n (p)|^2  
\end{equation}
for measurements of the momentum $P$ to yield a result in $\Delta\subset \mathcal M$.

It is invertibly related to the more abstract functions $\Psi\in \mathcal H_R$, which have a fixed dependence 
$\Psi(gh)=R^{-1}(h)\Psi(g)$ on right factors $h\in H$, by a section $\sigma$ of the bundle $G$ over the base~$\mathcal M$, in physics parlance by a choice of a gauge.
Mathematically~$\sigma$ is a set of smooth maps 
\begin{equation}
\sigma_\alpha: \mathcal U_\alpha \rightarrow G\ ,
\end{equation}
the local sections, with domains $\mathcal U_\alpha\subset \mathcal M$ which cover 
$\mathcal M = \cup_\alpha \mathcal U_\alpha$.
Each $\sigma_\alpha$ has to cut
each fiber over $\mathcal U_\alpha$ once: it chooses for each $p\in \mathcal U_\alpha$ a Lorentz transformation $\sigma_\alpha(p)$
which maps $\underline p$ to $p$. This relates in $\mathcal U_\alpha$ each function $\Psi\in \mathcal H_R$ to its momentum wave function 
$\psi_\alpha:\mathcal U_\alpha \rightarrow \mathbb C^{2s+1}$, 
\begin{equation}
\label{psimass}
\psi_\alpha = \Psi\circ \sigma_\alpha\ .
\end{equation}
In their common domain two local sections differ by their \idx{transition function} 
\begin{equation}
\label{transition}
\begin{gathered}
h_{\alpha\beta}: \mathcal U_\alpha\cap \mathcal U_\beta \rightarrow H\ ,\\  
\sigma_\beta(p) = \sigma_{\alpha}(p)\,h_{\alpha\beta }(p)\quad \text{(no sum over $\alpha$)}\ ,\ 
h_{\alpha\beta}^{-1}(p)= h_{\beta\alpha}(p)\ .
\end{gathered}
\end{equation}
As $\Psi$ is in $\mathcal H_R$, momentum wavefunctions are related in $ \mathcal U_\alpha\cap \mathcal U_\beta$ by
\begin{equation}
\label{uebergang}
\psi_\beta = R(h_{\beta \alpha}) \psi_\alpha\quad  \text{(no sum over $\alpha$)}\ .
\end{equation}
The other way round, to each set of functions $\psi_\alpha:\mathcal U_\alpha \rightarrow \mathcal V$
with such transition functions there corresponds the function $\Psi\in \mathcal H_R$
which for $g$ in a fiber over $\mathcal U_\alpha$ is defined by
\begin{equation}
\Psi(g) = R\bigl(g^{-1}\sigma_\alpha(g \underline p)\bigr)\,\psi_\alpha(g \underline p)\quad  \text{(no sum over $\alpha$)}\ . 
\end{equation}
In the fibers over $\mathcal U_\alpha\cap \mathcal U_\beta$ the local sections $\sigma_\alpha$ and~$\sigma_\beta$
yield the same $\Psi$.

Left multiplication by $g$ maps $\sigma_\alpha(p)$ to $g \sigma_\alpha(p)$
in the fiber over some domain $\mathcal U_\beta$ and 
related to $\sigma_\beta(g p)$ by an $H$-trans\-for\-ma\-tion, the \idx{Wigner rotation} 
\begin{equation}
\label{wignerab}
W_{\beta\alpha}(g,p)= \bigl(\sigma_\beta(g\,p)\bigr)^{-1} g\, \sigma_\alpha(p)\ .
\end{equation}

It enters as follows the transformation of the momentum wave functions,\quad (no sum over $\alpha$)
\begin{equation*}
\begin{gathered}
\bigl(U_g \psi\bigr)_\beta(g p)\stackrel{\ref{psimass}}{=}
\bigl(U_g \Psi\bigr)(\sigma_\beta(g p))\stackrel{\ref{wignerab}}{=}
\bigl(U_g \Psi\bigr)(g \sigma_\alpha(p) W_{\beta\alpha}^{-1})\stackrel{\ref{psifaser1}}{=}\\
R(W_{\beta\alpha})\bigl(U_g \Psi\bigr)(g \sigma_\alpha(p))\stackrel{\ref{psibuendel}}{=}
R(W_{\beta\alpha})\, \Psi(\sigma_\alpha(p))\stackrel{\ref{psimass}}{=}
R(W_{\beta\alpha} )\, \psi_\alpha(p)\ ,
\end{gathered}
\end{equation*}
\begin{equation}
\label{lorentzrep}
\bigl(U_g \psi\bigr)_\beta(g p) = R\bigl(W_{\beta\alpha}(g,p)\bigr)\, \psi_\alpha(p)\ .
\end{equation}
These transformations of momentum wave functions represent~$G$, as
for $p\in \mathcal U_\alpha$, $g_1 p\in \mathcal U_\beta$ and $g_2g_1 p\in \mathcal U_\gamma$
the Wigner rotations satisfy  by (\ref{wignerab})
\begin{equation}
\label{indmult}
W_{\gamma\alpha}(g_2g_1,p)= W_{\gamma\beta}(g_2,g_1\,p)\,W_{\beta\alpha}(g_1, p)\quad  \text{(no sum over $\beta$)}\ .
\end{equation}
One can drop the denomination of the neighbourhoods in (\ref{lorentzrep})
\begin{equation}
\bigl(U_g \psi\bigr)(g p) =  
R(W(g,p))\, \psi(p)\ ,
\end{equation}
if one agrees to read equations about values of sections 
to apply by definition in the neighbourhoods which contain the arguments.

In $D=4$ for massless representations with non-trivial~$R$ the complication of several local sections is unavoidable.
The Lorentz group $G$ is a non-trivial bundle over $\mathcal M_0=\set{p: p = \e^a (1,\vec n),\, a \in \mathbb R,\, \vec n \in S^{2}} $ 
which has the non-trivial topology of $\mathbb R\times S^2$. The subgroup SU$(2)\sim S^3 \subset G$
is a bundle over $S^2\subset \mathcal M_0$ with fibers diffeomorphic to $S^1$, each winding once around each other fiber.
\begin{figure}[h] 
\label{hopfbundel}
\centering
\includegraphics[scale=0.66]{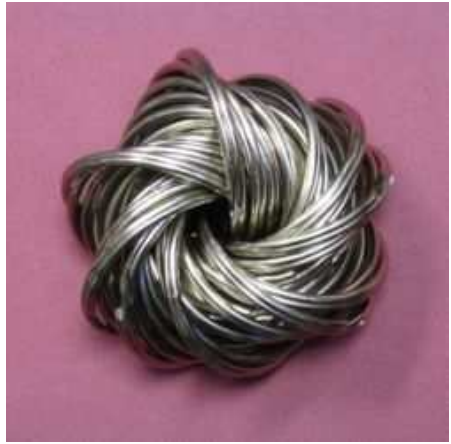}
\caption{Stereographic Projection of the Hopf Fibration of $S^3$ \cite{richter}}
\end{figure}
This bundle, the Hopf bundle, does not allow to extend a local section smoothly to a global section over $\mathcal M_0$.
This is demonstrated by the interlocked fibers, which one can cut once -- all but one -- in their intersection with a disk bordered by one chosen 
$S^1$.  But this chosen $S^1$ is not cut in exactly one point. 
Mathematically: if a global section existed, then $G$ would be a  product manifold $\mathcal M_0 \times H $ of the base and  a fiber.
\footnote{As Florian Oppermann pointed out to me
SO$(1,D-1)\sim$ SO$(D-1)\times \mathbb R^{D-1}$ 
factorizes  into \\ $\mathcal M_0\times E(D-2)\sim S^{D-2}\times $SO$(D-2)\times \mathbb R^{D-1}$ 
only in the 4 cases $D-1=1,2,4,8$  \cite[page 292]{hatcher}.} 
But $S^3\subset G$ is simply connected -- each closed path in $S^3$ can be continuously shrunk to a point -- 
while  $S^1$ is not, hence $S^3\ne S^2\times S^1$. 
So massless representations with nontrivial $R$ have to employ at least two local sections.


\section{Smooth Wavefunctions}

Each $\omega$  from the Lie algebra~\idx{$\mathfrak g$} of a finite dimensional Lie group $G$
generates by the exponential map the elements $g_t=\e^{t\,\omega}$, $t\in \mathbb R$,
of a one-parameter group. 
The skew hermitian generators $-\ir M_\omega$ of the one-parameter subgroups of a unitary representation $g \mapsto U_g$ , 
$U_g:\mathcal H(\mathcal M)\rightarrow \mathcal H(\mathcal M)$, 
represent $\mathfrak g$ on the subspace of states $\Psi$ on which all $U_{\e^{t\omega}}$ act smoothly \cite{schmuedgen}
\begin{equation}
-\ir M_\omega \Psi = \lim_{t\rightarrow 0} (U_{\e^{t\omega}}\Psi - \Psi) /t\ .
\end{equation}
They generate $U_{\e^{t\,\omega}}=\e^{-\ir\, t M_\omega}$ by their ($\omega$-dependent) spectral resolution~$E_\lambda$ 
\begin{equation}
M_\omega = \int\! \dv\! E_{\lambda}\,\lambda\ ,\ 
U_{\e^{t\,\omega}} = \e^{-\ir\,t\, M_\omega} = \int\! \dv\! E_{\lambda}\,  \e^{-\ir\,  t\,\lambda}\,\ .
\end{equation}
The unitary operators $U_{\e^{t\,\omega}}$ are defined in the complete Hilbert space $\mathcal H$,  while the power series $\sum_k (-\ir t M_\omega )^k \Psi/k!$ converges only 
in an analytic subspace which is unsuitably small for some purposes, e.g. it contains no state~$\Psi$
of compact support.

Because the maps $g \mapsto U_g \Psi$  are measurable therefore the smoothened states
\begin{equation}
\Psi_f = \int_{\mathfrak P}\! \dv \mu_g\, f(g)\, U_g \Psi
\end{equation}
exist, which are averaged with smooth functions $f: \mathfrak P \rightarrow \mathbb C$ of \emph{compact} support
and with an invariant volume form $\dv  \mu_g$, $\dv \mu_{g' g}=\dv  \mu_g$. These states transform by
\begin{equation}
U_g \Psi_f = \Psi_{f\circ g^{-1}}\ .
\end{equation}
Consequently they are infinitely differentiable with respect to the coordinates~$\omega$.
They span the \index{Garding@G\aa rding space}G\aa rding space, which is dense in $\mathcal H $ and invariant under all $U_g$ and 
their generators. It is the maximal common domain  of the algebra of the skew hermitian generators
\cite{dixmier, schmuedgen}.

That  such a common and dense domain exists for each unitary, strongly measurable\index{strong limit} representation of a finite-dimensional Lie group
justifies in retrospect physicists who manipulate the unbounded generators algebraically not caring about domains. 

The rough which one has to take with the smooth: an operator can be composed with a generator of a Lie group only if it maps smooth states to smooth states.
Otherwise the algebraic product generator times operator is not defined.
Rough operators with discontinuities or singularities in the group orbit,
and be it only in a single point, cannot occur in the algebra of
the Poincar\'e generators.  
Rough operators map physical states to unphysical states on which the Poincar\'e generators diverge.

\section{Massive Representations}

If one picks a momentum $\underline p$ from a massive shell $\mathcal M_m$, $m > 0$ (\ref{massshell}),
then for an observer at relative rest it has the coordinates $\underline p = (m,0,\dots)$
and $\mathcal M_m$ is its orbit under the restricted Lorentz group. 
The stabilizer of $\underline p$ is the group of rotations, SO$(D-1)$ which in $D=4$ is covered by $H=$SU$(2)$.
Each of its irreducible representations $R$ is determined by its spin~$s$ 
where $2s+1$, the dimension of~$R$, is a natural number.

Each Lorentztransformation $\Lambda=L_u O$ can be uniquely and continuously decomposed into 
a rotation $O$, $O^{\T}=O^{-1}\!,$ and  a boost $L_u=L_u^{\T}$ \cite{dragon1},
\begin{equation}
\label{lorp}
L_u = \begin{pmatrix}
\sqrt{1+\vec u^2}&  \vec  u^{\T}\\
 \vec   u &\quad  \eins+ \frac{ \vec  u\, \vec  u^{\T}}{1+\sqrt{1+\vec u^2}}
\end{pmatrix}\ .
\end{equation}
The boost $L_u$ maps the four-momentum $\underline{p}=(m,0,0,0)$ of a massive particle at rest
to the momentum $p=(\sqrt{m^2+\vec p^2},\vec p)= m\, u$ 
of a particle with four-velocity $u$. 

It provides the global section  
\begin{equation}
\sigma:\mathcal M_m \rightarrow \text{SO}(1,D-1)^\uparrow\ ,\ 
\sigma(p) = L_{p/m}\ ,\  L_{p/m}\, \underline p = p\ .
\end{equation}

This choice of a section allows to determine explicitly the generators of (\ref{lorentzrep}) by differentiation  with respect to $t$ at $t=0$.
The left side  yields $\omega^i(p)\, \partial_i \Psi -\ir\, M_\omega\,\Psi$.
\footnote{The  summation index $i$ enumerates coordinates of the mass shell, in terms of which Lorentz transformations
act smoothly e.g. the spatial components of the momentum.
The derivatives~$\partial_{p^n},\ n\in \set{0,1,2,3}$, however, are not defined on wave functions of $\mathcal M_m$. 
They are \emph{not} functions of $\mathbb R^{1,3}$.
}

Rotations $D=\e^\omega$ rotate four-velocities, $D L_u D^{-1}= L_{Du}$, thus they agree with their Wigner
rotation, $W(D,p)=L_{Du}^{-1}D L_{u}= D$.
Let $D_t=\e^{t\omega}$ be unitarily represented by 
$R(\e^{t\omega})\Psi(p) =\bigl(\exp \frac{t}{2}\omega^{ij}\Gamma_{ij}\bigr)\Psi(p)$ with skew hermitian 
matrices $\Gamma_{ij}$, which represent $\mathfrak{so}(3)$ or $\mathfrak{su}(2)$
($i,j,k,l\in\Set{ 1,2,3}$)
\begin{equation}
\label{drehalgebra}
 [\Gamma_{ij},\Gamma_{kl}]= \delta_{ik}\Gamma_{jl}- \delta_{il}\Gamma_{jk} - \delta_{jk}\Gamma_{il}+ \delta_{kl}\Gamma_{ik}\ ,
\ \Gamma_{ij} = - \Gamma_{ij}^{\,*\,\T} = - \Gamma_{ji}\ .
\end{equation}
The derivative of $R(\e^{t\omega})$ with respect to $t$ at $t=0$ is $\frac{1}{2}\omega^{ij}\Gamma_{ij}\Psi(p) $. 
Hence (\ref{lorentzrep}) implies
\begin{equation}
\label{unitdreh}
\bigl(-\ir M_{ij}\Psi\bigr)(p) = -\bigl(p^i \partial_{p^j} - p^j \partial_{p^i}\bigr)\Psi(p) + \Gamma_{ij}\Psi(p)
\ .
\end{equation}
As $-\ir M_{ij}$ generate rotations, 
    $(J^1,J^2,J^3)=(M_{23}, M_{31},M_{12})$ are the components of the \idx{angular momentum} operator $\vec J$.
It consists of orbital angular momentum \mbox{$\vec L = -\ir p \times \partial_p$}  and spin $\vec S$
contributed by the matrices~$\ir \Gamma$, $\vec J = \vec L + \vec S$,  
$[L^i,S^j]=0$.

The component functions of the differential operator on the right side of (\ref{unitdreh}) are the \emph{negative} \label{negcoeff} of the 
infinitesimal motion $\delta p$ of the points $p$.
This motion occurs on the left side of (\ref{lorentzrep}) or 
as inverse transformation $M_g^{-1}$ in the adjoint transformation
\begin{equation}
\Ad_g(f) = N_g \circ f \circ M_g^{-1}
\end{equation}
of maps $f:\mathcal M \rightarrow \mathcal N$ of manifolds $\mathcal M$ to $\mathcal N$ on which $G$ 
is realized by transformations $M_g$ and $N_g$, $M_g M_{g'}= M_{g g'}$, $N_g N_{g'}= N_{g g'}$, $\Ad_g \Ad_{g'}= \Ad_{g g'}$.

The Wigner rotation $W(L_u,p)$ rotates from $p$ to~$L_u p$ 
by the deficit angle~$\delta$ of the hyperbolic triangle with vertices $(L_u p, p, \underline p)$ 
and lengths $a=\dist{(\underline p, p)}$ and $b=\dist{(L_u p,p))}$ and $\gamma$ the angle included by $\vec p$ and $\vec u$
\begin{equation}
\tan \frac{\delta}{2}= \frac{\sin \gamma}{\bigl(\coth \frac{a}{2}\,\coth \frac{b}{2}\bigr)+\cos\gamma} 
\ .
\end{equation}
We derive this result (\ref{wignerdelta}) in appendix A. Crucial as it is for the determination of the generators, I am unaware of
a published derivation. For infinitesimal $L_u$ and $b$ this yields
\begin{equation}
\frac{\dv \delta}{\dv b}_{|_{b=0}}= (\sin \gamma) \tanh \frac{a}{2}= (\sin \gamma) \frac{\sh a}{\ch a + 1}=
(\sin \gamma) \frac{|\vec p|}{p^0 + m}\ .
\end{equation}
As the Wigner rotation of an infinitesimal boost $l_{0i}$ applied to $p$ rotates from $\vec p$ to $- e_i$ 
or from $ e_i$ to $\vec p$ it is represented by
$\Gamma_{ij}{p^j}/{(p^0 + m)}$.
Combining both infinitesimal changes, we obtain the generators of boosts 
\begin{equation}
\label{unitboost}
\bigl(-\ir M_{0i}\Psi\bigr)(p)= p^0\,\partial_{p^i}\Psi(p) + \Gamma_{ij}\frac{p^j}{p^0 + m}\Psi(p)\ .
\end{equation}

\longpage

The generators (\ref{unitdreh}, \ref{unitboost}) are skew hermitian with respect to the scalar pro\-duct 
(\ref{lorentzmaas}).
They rotate and boost the four-vector $P$
\begin{equation}
\label{fourvec}
[-\ir M_{mn}, P^r] = - \delta_m{}^r\,P_n + \delta_n{}^r\,P_m\ ,
\end{equation}
and represent the Lorentz algebra (\ref{loralgebra}).
For $[-\ir M_{ij},-\ir M_{kl}]$ this is simple to check. It is manifest for $[-\ir M_{ij},-\ir M_{0k}]$, because
$p^i, \partial_{p^i}$ and $\Gamma_{ij}$ transform as vectors or pro\-ducts of vectors under the 
rotations of momentum and spin. The relations \mbox{$[-\ir M_{0i},-\ir M_{0j}]=\ir M_{ij}$} 
hold as if by miracle\footnote{Observe the summation range $p^kp^k=(p^0)^2-m^2=(p^0+m)(p^0-m)$\,.} 
\begin{align}
\nonumber
\  [p^0\partial_{p^i}&+ \Gamma_{ik}\frac{p^k}{p^0 + m}\,,\, p^0\partial_{p^j} + \Gamma_{jl}\frac{p^l}{p^0 + m}]=\\
\nonumber &= \Bigl(p^0 \frac{p^i}{p^0}\partial_{p^j} + \frac{1}{(p^0 + m)^2}(p^0(p^0 + m) \delta_i{}^l-p^ip^l)\Gamma_{jl}\Bigr)
- \Bigl(i\leftrightarrow j\Bigr)+\\
\nonumber &\quad +\frac{p^kp^l}{(p^0 + m)^2}\bigl(\delta_{ij}\Gamma_{kl}-\delta_{il} \Gamma_{kj}-\delta_{kj} \Gamma_{il}+\delta_{kl} \Gamma_{ij}\bigr)\\
&= - \bigl(-(p^i\partial_{p^j}-p^j\partial_{p^i})+ \Gamma_{ij}\bigr)\ .
\label{miracle}
\end{align}

Splitting the momenta  $p = m u$ of the wavefunctions into the mass $m$ and the four-velocity~$u=(\sqrt{1+\vec u^2},\vec u)$  and allowing the  mass $m$ 
to depend on further invariant variables~$r$, as is the case for many-particle states, 
the generators of \emph{each} massive spin-$s$ representation, also each interacting one, are unitarily equivalent to
\begin{equation}
\begin{gathered}
P^m = U^m\, M\ ,\  0=[M, U^m] =  [M, M_{mn}]=[U^m,U^n]\ ,\ U^2 = 1\ , \\ \bigl(U^m \Psi\bigr)(u,r) = u^m\, \Psi(u,r)\ ,\  \bigl(M \Psi\bigr)(u,r) = m(r) \, \Psi(u,r)\ ,\\
\label{gencont}
\begin{aligned}
\bigl(-\ir M_{ij}\Psi\bigr)(u,r) &= -\bigl(u^i \partial_{u^j} - u^j \partial_{u^i}\bigr)\Psi(u,r) + \Gamma_{ij}\Psi(u,r)\ ,\\
\bigl(-\ir M_{0i}\Psi\bigr)(u,r) &= \sqrt{1+\vec u^2}\,\partial_{u^i}\Psi(u,r) + \Gamma_{ij}\frac{u^j}{1+\sqrt{1 +\vec u^2}}\Psi(u,r)\ .
\end{aligned}
\end{gathered}
\end{equation}

For fixed $r$, the operators $U^m$ and $M_{mn}$ generate the irreducible unit mass spin-$s$ representation of $\mathfrak P$ in $\hat{\mathcal H_s}$ (\ref{hred}).
The space of one-particle states is the  span of the eigenspaces of $M$.

Also each massive interacting representation has this form.
Its mass $M'$ commutes with all generators~$U^m$ and $M_{mn}$ and is therefore invariant under translations $V_a = \e^{\ir U \cdot a}$, generated by $U^m$
and under Lorentztransformations. To yield a non-trivial $S$-matrix however, it must not commute with $M$ \cite{dragon3}.
Both $M$ and $M'$ are positive operators in the invariant Hilbert space $\mathcal I_s$ of wavefunctions of $r$.

On many-particle states, $U^m$ generates translations of the center while $M'$ generates the interacting relative motion. The Hamiltonian $P^{\prime 0}= U^0 M'$
separates not as a sum of Hamiltonians of the motion of the center and of the interacting relative motion, but factorizes as their product \cite{dragon3}.


\section{Massless Representations and Helicity}

Picking arbitrarily from the massless shell 
\begin{equation}
\mathcal M_0 = \Set{(p^0,\vec p) : p^0 = |\vec p|> 0\ ,\ \vec p\ne 0 }\ ,
\end{equation}
a momentum~$\underline p$, it has for a suitable observer coordinates $\underline p = (1,0,0,\dots 1)$
and~$\mathcal M_0$ is its Lorentz orbit. It is the manifold $S^{D-2}\times \mathbb R$,
as $\vec p\ne 0$ is specified by its direction and its nonvanishing modulus $|\vec p| = \e^a> 0$. 
Though the observer has adopted a unit of energy, this does not spoil dilational symmetry
as dilations map $\mathcal M_0$ to itself and the transformations~$U_{\e^\lambda}$,
\begin{equation}
(U_{\e^\lambda} \Psi)(p) = \e^{-\lambda}\,\Psi(\e^{-\lambda} p)\ ,
\end{equation}
represent them unitarily with respect to the scalar product (\ref{lorentzmaas}) with $m=0$.

The stabilizer~$H$ of $\underline p$ is generated by infinitesimal Lorentz transformations~$\omega$,
$\eta \omega =-(\eta \omega)^{\text{T}}$, with $\omega \underline p = 0$, $\omega^m{}_0 + \omega^m{}_z=0$, 
thus of the form
\begin{equation}
\omega(\vec a,\hat{\omega})=
\begin{pmatrix}
   &  \vec a^{\text{T}} &\\
\vec a  & \hat\omega^{\phantom{T}}    & -\vec a\!\\
   & \vec a^{\text{T}}  &\\
\end{pmatrix}\ .
\end{equation}
Here $\hat\omega$ is a skewsymmetric $(D-2)\times (D-2)$ matrix,
which generates a rotation $w\in\,$SO$(D-2)$ and $\vec a$ is a $D-2$-vector.
Because $\omega(\vec a,0)$ and $\omega(\vec b,0)$ 
commute they generate translations in  $D-2$ dimensions. They are rotated by $w$. 
So $H$ is (the cover of) the Euclidean group \idx{$\text{E}(D-2)$} 
in  $\mathbb R^{D-2}$.

By Mackey's theorems \cite{mackey}  each massless unitary representation  of the Poincaré group $\mathfrak P$
is a sum of irreducible representations which are induced by unitary irreducible representations of E$(D-2)$. 

By the same theorems each such representation of $E(D-2)$ acts on the functions of the 
SO$(D-2)$ orbit, $S^{D-3}$ or $\Set{0}$,  of some $\underline q\in \mathbb R^{D-2}$
and is characterized by a unitary irreducible representation of $q$'s stabilizer SO$(D-3)$ or, if $q=0$, SO$(D-2)$. 

\label{continuousspin}
In $D=4$, the  SO$(2)$-orbit of $\underline q\ne 0$ is a circle, its stabilizer is trivial. The induced unitary representation of E$(2)$ acts
on wave functions  of a circle. Contrary to its denomination '\idx{continuous spin}' such a representation contains only integer or half integer 
(infinitely many) helicities, the Fourier modes of the functions of the circle. Such a representation is not contained in the restriction
of a finite-dimensional representation of the Lorentz group to E$(2)$ but requires a field which in addition depends
on additional continuous variables \cite{mund}. As these infinitely many massless states per given four-momentum 
make the specific heat of each cavity infinite \cite[p. 684]{wigner1948} they have to decouple from ordinary matter 
early enough after the hot big bang so as not to  spoil our understanding of the thermodynamic evolution of the universe and its observed remnant, the microwave background radiation.\index{background radiation}

If $\underline q=0\in \mathbb R^2$, then its orbit under SO$(2)=E(2)/\mathbb R^2$ only consists of $\underline q$. All translations $T_{\vec a}$ in E$(2)$ are represented 
trivially by $T_{\vec a}=\e^{\ir\, \underline q \cdot a}=\e^0 = 1$. Each unitary irreducible representation $R$ of SO$(2)$ is one-dimensional
and represents a rotation $D_\delta$ by the angle $\delta$ by multiplication with 
\begin{equation}
R_{D_\delta}=\e^{-\ir h \delta}
\end{equation}
where in ray representations $h$, the helicity, is some real number.

In $D=4$ there is no global section 
with which to relate functions in $\mathcal H_R$ (\ref{psifaser1}) to momentum wave 
functions of~$\mathcal M_0$ (\ref{psimass}). 
Therefore we use two local sections $N_p$  and $S_p$  to derive the transformation of momentum wave functions.
They are differentiable in the north $\mathcal U_N$ outside the negative $3$-axis 
$\mathcal A_-$
 \index{$\mathcal U_N, \mathcal U_S, \mathcal A_-,\mathcal A_+$}
and in the south $\mathcal U_S$ outside the positive $3$-axis $\mathcal A_+$
\begin{align}
\nonumber
\mathcal U_N &= \Set{p \in \mathcal M_0: |\vec p| + p_z > 0}\ , &
\mathcal U_S &= \Set{p \in \mathcal M_0: |\vec p| - p_z > 0}\ ,\\
\label{nsumg}
\mathcal A_-&=\Set{ (\lambda,0,0,-\lambda): \lambda > 0 }\ ,&
\mathcal A_+&=\Set{ (\lambda,0,0,\lambda): \lambda > 0 }\ .
\end{align}
As $\mathcal U_N$ and also $\mathcal U_S$ cover $\mathcal M_0$ up to a set of vanishing measure, one need not distinguish integrals
on these sets though as domains of differentiable functions one has to discriminate these sets carefully.

The northern section \label{northernsection} $N_p=D_p B_p$ boosts $\underline p = (1,0,0, 1)$ 
in $3$-direction to $B_p \underline p = (|\vec p|,0,0, |\vec p|)$
and rotates  around $(-\sin \varphi, \cos\varphi,0)$ 
by~$\theta$, $0\le \theta < \pi$, 
to  $p = |\vec p| (1,\sin\theta\cos \varphi, \sin\theta\sin\varphi,\cos \theta)= |\vec p|(1,n_x,n_y,n_z)$,
\begin{equation}
\label{bp}
\begin{aligned}
D_p &= 
\begin{pmatrix}
c_{\varphi} & -s_{\varphi} &\\
s_{\varphi} & \phantom{-}c_{\varphi}& \\
& & 1
\end{pmatrix}
\begin{pmatrix}
\phantom{-}c_{\theta} &  & s_{\theta}\\
& 1 & \\
-s_{\theta}& & c_{\theta}
\end{pmatrix}
\begin{pmatrix}
\phantom{-}c_{\varphi} & s_{\varphi} &\\
-s_{\varphi} & c_{\varphi}& \\
& & 1
\end{pmatrix}\\
&= 
\begin{pmatrix}
1  - \frac{n_x^2}{1 + n_z}& -\frac{n_xn_y}{1 + n_z}& n_x\\
-\frac{n_xn_y}{1 + n_z}   & \phantom{-}1 - \frac{n_y^2}{1 + n_z}& n_y\\
-n_x & -n_y & n_z
\end{pmatrix}\ .
\end{aligned}
\end{equation}

For $D_p$ to be smooth at $\underline p$ it is \emph{crucial} that, different from the prevalent literature \cite[volume I, (2.5.47)]{weinberg}, $D_p$
acts first by the inverse rotation by $\varphi$ around $e_3$ and not only by the two rotations
by the Euler angles $\theta$ around $e_2$ and $\varphi$ around~$e_3$. Their product is discontinuous at $\theta=0$.

In place of the unhandy $4\times 4$ Lorentz matrix $N_p$ one works more easily with
its SL$(2,\mathbb C)$ representation $n_p$ (\ref{uvonalpha}, \ref{eh}) using $p^\prime = \sqrt{|\vec p|}$,  
$\theta^\prime = {\theta}/{2}$ and
\begin{equation}
\tan \theta'=
\frac{2 \cos\frac{\theta}{2}\sin\frac{\theta}{2}}{2\cos^2\frac{\theta}{2}}=
\frac{\sin\theta}{1+\cos\theta} =
\frac{\sqrt{p_x^2+p_y^2}}{|\vec p| + p_z}= \sqrt{\frac{|\vec p|-p_z}{|\vec p|+p_z}}\ ,
\end{equation}
\begin{equation}
\label{nsection}
n_p = 
\begin{pmatrix}
\frac{\cos \theta^\prime}{p^\prime} & -p^\prime \sin \theta^\prime\e^{-\ir \varphi}\\
 \frac{\sin \theta^\prime}{p^\prime}\, \e^{\ir \varphi} & p^\prime \cos \theta^\prime\\
\end{pmatrix}
=
\frac{1}{\sqrt{2}}
\begin{pmatrix}
\frac{\sqrt{|\vec p|+p_z}}{|\vec p|} & -\frac{p_x-\ir p_y}{\sqrt{|\vec p|+p_z}}\\
\frac{p_x+\ir p_y}{|\vec p|\sqrt{|\vec p|+p_z}} & \vphantom{T^{T^{T}}}\sqrt{\scriptstyle |\vec p|+p_z}
\end{pmatrix}
\ .
\end{equation}

Check that $n_p$ transforms $\hat{\underline p}=(\eins- \sigma^3)$
to $n_p \,\hat{\underline p}\,(n_p)^\dagger= |\vec p| - \vec p\cdot \sigma=\hat p$ 
and for $b_p=1$
leaves invariant the axis $-p_y \sigma_x + p_x \sigma_y $.
In $\mathcal U_N$, $0\le \theta < \pi$, the section $n_p$ is smooth.
It is discontinuous at~$\mathcal A_-$
where the limit $\theta \rightarrow \pi$  depends on~$\varphi$. 

The section
\begin{equation}
\label{suednordsl2}
s_p = 
\begin{pmatrix}
\frac{\cos \theta^\prime}{p^\prime}\e^{-\ir \varphi} & -p^\prime \sin \theta^\prime\\
 \frac{\sin \theta^\prime}{p^\prime}  & p^\prime \cos \theta^\prime\e^{\ir \varphi}
\end{pmatrix}
=
\frac{1}{\sqrt{2}}
\begin{pmatrix}
\frac{p_x-\ir p_y}{|\vec p|\sqrt{|\vec p|-p_z}} & -\sqrt{\scriptstyle |\vec p|-p_z}\\
\frac{\vphantom{T^{T^{T^T}}}\sqrt{|\vec p|-p_z}}{|\vec p|} & \frac{p_x+\ir p_y}{\sqrt{|\vec p|-p_z}}
\end{pmatrix}
\end{equation}
is defined and smooth in the south, $0<\theta \le \pi$. The corresponding
southern section~$S_p$  
rotates $B_p \underline p$ in the $1$-$3$-plane by $\pi$ to $|\vec p|(1,0,0,-1)$
and then along a great circle by the smallest angle to~$p$, namely by $\pi - \theta$ around the axis
$(\sin \varphi, -\cos\varphi,0)$. 

In their common domain local SL$(2,\mathbb C)$ sections differ
by the multiplication from the right (\ref{transition}) with a momentum dependent
SL$(2,\mathbb C)$ matrix which represent E$(2)$
\begin{equation}
\label{hgruppe}
w=
\begin{pmatrix}
\e^{-\ir \delta/2} & 0 \\
-a & \e^{\ir \delta/2}\\
\end{pmatrix}\ .
\end{equation}
In the case at hand, $s_p$ differs in $\mathcal U_N\cap \mathcal U_S$
from $n_p$ by a preceding rotation around the $3$-axis by $2\varphi(p)$. Geometrically this angle is the area of the spherical lune
(spherical digon) with  vertices $\vec e_z$ and $- \vec e_z$ and sides (meridians) through~$\vec e_x$ and $\vec p/|\vec p|$,
\begin{equation}
s_p =  
n_p
\begin{pmatrix}
\e^{-\ir \varphi(p)}& \\
& \e^{\ir \varphi(p)}\!
\end{pmatrix}
\ ,\ 
\label{varphi}
\e^{\ir \varphi(p) }= 
\frac{p_x+\ir\,p_y}{\sqrt{(|\vec p|-p_z)(|\vec p|+p_z)}}\ .
\end{equation}

The angle $\delta$ of the Wigner rotation $W(\Lambda, p)$ 
$\Lambda N_p = N_{\Lambda p}W(\Lambda,p)$ (\ref{wignerab}), 
can be easily read off the Iwasawa decomposition (\ref{e2transz}) of the SL$(2,\mathbb C)$ representation 
\begin{equation}
\label{wignerhel}
\lambda\, n_p = n_{\Lambda p}\, w(\Lambda, p)
\end{equation}
as twice the phase of $(n_{\Lambda p}\,w)_{22} =|(n_{\Lambda p}\,w)_{22}|\, \e^{\ir\delta/2}$.

If $\Lambda$ is a rotation by $\alpha$ around an axis $\vec n$ then $\lambda$ 
is given by  (\ref{uvonalpha}) and  the Wigner angle $\delta$ and its infinitesimal value are
\begin{equation}
\label{deltainf}
\tan \frac{\delta}{2} = 
\frac{n_z |\vec p|+\vec n\cdot\vec p}{(|\vec p|+p_z)(\cot \frac{\alpha}{2})+ n_xp_y-n_yp_x}\ ,\ 
\frac{\dv \delta}{\dv \alpha}_{|_{\alpha=0}}= 
 \frac{n_z |\vec p| + \vec n \cdot \vec p}{|\vec p| + p_z}\ .
\end{equation}

As $R_{D_\delta} = \e^{-\ir h \delta}$, the representation  $-\ir\, h\,p_y/(|\vec p| + p_z)$ of the infinitesimal Wigner rotation
accompanies e.g. the infinitesimal rotation in the $3$-$1$-plane. 

Analogously one determines the Wigner angle of a boost (\ref{eh})
\begin{equation}
\tan \frac{\delta}{2} = 
\frac{-(n_xp_y-n_yp_x)}{(|\vec p|+p_z)(\coth \frac{\beta}{2})+ n_z |\vec p| +  \vec n\cdot \vec p}\ ,\ 
\frac{\dv \delta}{\dv \beta}_{|_{\beta=0}}= 
 \frac{-(n_xp_y-n_yp_x)}{|\vec p| + p_z}
\end{equation}
and, as $-\ir M_{0i} $ boosts from $e_0$ to~$-e_i$, obtains in $D=4$ 
\cite{bose, ek, lomont}
\begin{equation}
\label{masselosnord}
\begin{aligned}
\bigl(-\ir M_{12}\Psi\bigr)_N(p)&= - \bigl(p_x\partial_{p_y} - p_y\partial_{p_x}\bigr)\Psi_N(p) - \ir\, h\,\Psi_N(p)\ ,\\
\bigl(-\ir M_{31}\Psi\bigr)_N(p)&= - \bigl(p_z\partial_{p_x} - p_x\partial_{p_z}\bigr)\Psi_N(p) - \ir\, h\,\frac{p_y}{|\vec p| + p_z}\Psi_N(p)\ ,\\
\bigl(-\ir M_{32}\Psi\bigr)_N(p)&= - \bigl(p_z\partial_{p_y} - p_y\partial_{p_z}\bigr)\Psi_N(p) + \ir\, h\,\frac{p_x}{|\vec p| + p_z}\Psi_N(p)\ ,\\
\bigl(-\ir M_{01}\Psi\bigr)_N(p)&=  |\vec p| \partial_{p_x} \Psi_N(p) -\ir\,h\,\frac{p_y}{|\vec p| + p_z}\Psi_N(p)\ ,\\
\bigl(-\ir M_{02}\Psi\bigr)_N(p)&=  |\vec p| \partial_{p_y} \Psi_N(p) +\ir\,h\,\frac{p_x}{|\vec p| + p_z}\Psi_N(p)\ ,\\
\bigl(-\ir M_{03}\Psi\bigr)_N(p)&=  |\vec p| \partial_{p_z} \Psi_N(p)\ .
\end{aligned}
\end{equation}
In $D$-dimensional spacetime the generators are\footnote{Note the range of summation  
$p^k p^k = |\vec p|^2- p_z^{\,2}=(|\vec p|+ p_z)(|\vec p|- p_z)$.} 
\begin{equation} 
\begin{aligned}
\bigl(-\ir M_{ij}\Psi\bigr)_N(p)&= - \bigl(p^i\partial_{p^j} - p^j\partial_{p^i}\bigr)\Psi_N(p) + h_{ij}\,\Psi_N(p)\ ,\\
\bigl(-\ir M_{zi}\Psi\bigr)_N(p)&= - \bigl(p_z\partial_{p^i} - p^i\partial_{p_z}\bigr)\Psi_N(p) + h_{ik}\frac{p^k}{|\vec p|+p_z}\,\Psi_N(p)\ ,\\
\bigl(-\ir M_{0i}\Psi\bigr)_N(p)&=  |\vec p|\partial_{p^i}\Psi_N(p) + h_{ik}\frac{p^k}{|\vec p|+p_z}\,\Psi_N(p)\ ,\\
\bigl(-\ir M_{0z}\Psi\bigr)_N(p)&=  |\vec p|\partial_{p_z}\Psi_N(p)\ , 
\end{aligned}
\end{equation}
where $p_z = p^{D-1}$,  $i,j,k\in \set{1,\dots D-2}$ and $h_{ij}= - h_{ji}$, $h_{ij}^{*\T} = - h_{ij}$,
generate a representation of SO$(D-2)$ (\ref{drehalgebra}).

In $D=4$ one has $h_{ij}=-\ir h \varepsilon^{ij}$, where the real number $h$,  the \idx{helicity}, is 
the angular momentum $\vec p \cdot \vec J/\abs{\vec p}$ in the direction of 
the momentum~$\vec p\ne 0$, 
\begin{equation}
\label{helizit}
\bigl(({p_x}\,M_{23}+ {p_y}\,M_{31}+ {p_z} \,M_{12})\Psi\bigr)_N(p) = h\,|\vec p|\,\Psi_N(p)\ .
\end{equation}

The mere fact, that differential operators satisfy a Lie algebra on some space of functions does not make them generators of a 
representation of the corresponding group.
This is demonstrated by the operators $-\ir M_{mn}$ (\ref{masselosnord}). 
On differentiable functions of the northern coordinate patch $\mathcal U_N$ (\ref{nsumg})
they satisfy the Lorentz Lie algebra (\ref{loralgebra}) in $D=4$ \cite{bose, ek, lomont} no matter which 
real value the helicity $h$ has.  The Lorentz Lie algebra does not restrict $2h$ to be an integer.
The operators are formally skew hermitian with respect to the Lorentz invariant measure $\dt p$,  formally only, because
the singularities at $|\vec p|+ p_z = 0$ need closer investigation.

The operators (\ref{masselosnord}) cannot generate the Lorentz group because the domain $\mathcal U_N$ of the differentiable functions is too small:
Lorentz generators  act on smooth states, which have to be defined \emph{everywhere} on the massless shell~$\mathcal M_0$.
The group acts transitively and contains for each massless momentum $p$ a rotation
which maps it to  $\hat p \in \mathcal A_-=\set{p: p^0 = |\vec p | = - p_z > 0}$.

For $h\Psi_N(\hat p) \ne 0$ the functions  $h (|\vec p| + p_z)^{-1}\Psi_N$ are not defined on $\hat p \in \mathcal A_-$ and seem to
contradict (\ref{psibuendel}) by which smooth states in $\mathcal H_R$
are transformed to smooth states. 
One cannot require all wave functions $\Psi_N$ to vanish on~$\mathcal A_-$ because
such functions do not span a space which is mapped to itself by rotations.
One also cannot take comfort in the misleading argument \cite{fronsdal, sexl}
that the set of singular points of e.g. $M_{13}\Psi$ has measure zero: 
The argument is irrelevant as not the measure of the
singular set matters but the measure $\tilde \mu(\Gamma_c)$ of the sets 
$\Gamma_c = \set{p:|M_{13} \Psi(p)|^2 > c }$, where $\Psi$ is large.
If the limit $ c \rightarrow \infty $ of $c\,\tilde \mu(\Gamma_c)$ does not vanish 
then $M_{13}\Psi$ is not square integrable.

That a set of measure zero does not count in quantum physics holds for the elements of each equivalence class of a wave function but \emph{not} for
smooth states and their domain. Each smooth equivalence class contains one unique smooth function on which the generators are defined and act smoothly.

For negative $p_z$ and with $x = (p_x^2 + p_y^2)/p_z^2$ one has  
\begin{equation}
|\vec p| + p_z=|\vec p| - |p_z|= |p_z|(\sqrt{1+x}-1)\le |p_z|\, \frac{x}{2}
\end{equation}
because the concave function $x \mapsto \sqrt{1+x}$ is bounded by its tangent at $x=0$. 
So
\begin{equation}
\label{1dpplus}
\frac{1}{|\vec p| + p_z}  \ge 
 \frac{2|p_z|} {p_x^2+p_y^2}  
\end{equation}
diverges in a neighbourhood $\mathcal U$ of $\hat p \in \mathcal A_-$ at least like 
the inverse square of the axial distance $r=\sqrt{p_x^{2\vphantom{k}}+p_y^2}$.

\longpage

If $h\,\Psi_N$ does not vanish in a neighbourhood $\mathcal U$ of $\hat p \in \mathcal A_-$ then, for $M_{31}\Psi_N$ to exist, it must not be differentiable there. 
Otherwise the multiplicative term of  $M_{31}\Psi_N$ dominates near $\hat p$ where it scales as $ |p_z| / r$. 
Its squared modulus integrated over a sufficiently small $\mathcal U$ in cylindrical coordinates is bounded from below by a
positive number times an $r$-integral over $r / r^2 \dv\! r$ which diverges at the lower limit $r=0$. 
Hence the multiplicative term alone diverges.

Near $\mathcal A_-$ the derivative term $D\Psi_N= -p_z \partial_{p_x}\Psi_N$ in $M_{31}\Psi_N$ has to cancel 
the multiplicative singularity $M \Psi_N$ up to a function~$\chi$, which is smooth. 
This linear inhomogeneous condition $(D + M)\Psi_N=\chi$ is solved by variation of constants $\Psi_N = f \Psi_S$ where $f$ 
satisfies the two homogeneous conditions
\begin{equation}
|p_z|\bigl(\partial_{p_x} - 2 \ir h \frac{p_y}{p_x^2+p_y^2}\bigr)f= 0\ ,\ |p_z|\bigl(\partial_{p_y} + 2 \ir h \frac{p_x}{p_x^2+p_y^2}\bigr)f= 0\ ,
\end{equation}
for both $M_{31}\Psi_N$ and  $M_{32}\Psi_N$ to exist. 
They determine $f(p)=\e^{-2\ir h \varphi(p) }$ up to a 
factor. 

\longpage

The function $\Psi_S$ is smooth in the southern coordinate patch $\mathcal U_S$ (\ref{nsumg})
and related in $\mathcal U_N\cap \mathcal U_S$ by the transition function $f^{-1}=h_{SN}$ to $\Psi_N$
\begin{equation}
\label{suednordwell}
\Psi_S(p) = h_{SN}(p)\,\Psi_N(p)\ ,\ h_{SN}(p) = \e^{2\ir\, h\varphi(p)}=\Bigl (\frac{p_x+\ir p_y}{\sqrt{p_x^2+p_y^2}}\Bigr)^{2h}\ .
\end{equation}

Each state $
\Psi$ is given by local sections
$\Psi_N$ and $\Psi_S$  of a bundle over $S^2\times \mathbb R$ with transition function $h_{SN}$ 
which is defined and smooth in $\mathcal U_N\cap \mathcal U_S$
only if $2h$ is integer. This is why the helicity of a massless particle  is integer or half integer.
Multiplying (\ref{masselosnord}) with $h_{SN}$ one obtains from  (\ref{suednordwell}) 
\begin{equation}
\label{masselossued}
\begin{aligned}
\bigl(-\ir M_{12}\Psi\bigr)_S(p)&= - \bigl(p_x\partial_{p_y} - p_y\partial_{p_x}\bigr)\Psi_S(p) + \ir\, h\,\Psi_S(p)\ ,\\
\bigl(-\ir M_{31}\Psi\bigr)_S(p)&= - \bigl(p_z\partial_{p_x} - p_x\partial_{p_z}\bigr)\Psi_S(p) - \ir\, h\,\frac{p_y}{|\vec p| - p_z}\Psi_S(p)\ ,\\
\bigl(-\ir M_{32}\Psi\bigr)_S(p)&= - \bigl(p_z\partial_{p_y} - p_y\partial_{p_z}\bigr)\Psi_S(p) + \ir\, h\,\frac{p_x}{|\vec p| - p_z}\Psi_S(p)\ ,\\
\bigl(-\ir M_{01}\Psi\bigr)_S(p)&=  |\vec p| \partial_{p_x} \Psi_S(p) +\ir\,h\,\frac{p_y}{|\vec p| - p_z}\Psi_S(p)\ ,\\
\bigl(-\ir M_{02}\Psi\bigr)_S(p)&=  |\vec p| \partial_{p_y} \Psi_S(p) -\ir\,h\,\frac{p_x}{|\vec p| - p_z}\Psi_S(p)\ ,\\
\bigl(-\ir M_{03}\Psi\bigr)_S(p)&=  |\vec p| \partial_{p_z} \Psi_S(p)\ .
\end{aligned}
\end{equation}
The same generators result 
if, along the lines of the derivation of (\ref{masselosnord}), one reads the Wigner angle $\delta$ from 
the phase of $(\lambda s_p)_{12}=-|(s_{\Lambda p}w)_{12}|\e^{\ir \delta/2}$.
All $M_{mn}\Psi$ are square integrable, rapidly decreasing and smooth  
if $\Psi$ is. 


For all $\omega$ in the Lorentz Lie algebra  the operators $-\ir M_\omega = - \ir/2\, \omega^{mn}M_{mn}$ are by construction 
the derivatives of unitary one-parameter groups 
\begin{equation}
\label{sufficient}
-\ir M_\omega \bigl(U_{\e^{t\omega}} \Psi\bigr) = \partial_t \bigl(U_{\e^{t\omega}} \Psi\bigr)\ ,
\end{equation}
which act on the dense and invariant domain $\mathcal D$ of smooth states, where the transformations $U_{\e^{\omega}}$ 
and their products represent unitarily the Lorentz group. So $-\ir M_\omega$ not only satisfy the Lorentz algebra but they are
skew-adjoint (by Stone's theorem) and generate a unitary representation of the Lorentz group.

In $D>4$ dimensions 
\begin{equation}
\Psi_S(p)=R^{2}_p\Psi_N(p)
\end{equation}
where $R_p$ represents the rotation from $\vec p_{\T}/|\vec p_{\T}|$
to $\vec e_x$ which leaves vectors orthogonal to the $\vec e_x-\vec p_{\T}$-plane pointwise invariant.

For helicity $h\ne 0$ each continuous momentum wave function $\Psi$ has to vanish along some line, 
a Dirac string in momentum space, from $\ln |\vec p|=-\infty $ to $\ln |\vec p|=\infty $. 
Namely, if one removes the set $\mathcal N$, where $\Psi$
vanishes, from the domains $\mathcal U_N$ and $\mathcal U_S$ then the remaining sets
$\hat{\mathcal U}_N$ and $\hat{\mathcal U}_S$ cannot both be simply connected.
In $\hat{\mathcal U}_N $ the phase of $\Psi_N$ is  continuous and
its winding number along a closed path, being integer, does not change under deformations of the path. 
For a contractible path this winding number vanishes, as the phase along the path becomes constant upon its contraction.
If $\hat{\mathcal U}_N $ is simply connected then it contains a contractible path around $\mathcal A_+$ which
also lies in $\hat{\mathcal U}_S$. 
On this path the phase of $\Psi_N$ has vanishing winding number but the phase of  $\Psi_S(p)=\e^{2\ir\,h\,\varphi(p)}\Psi_N(p)$ 
winds $2h$-fold. So the path cannot be contractible in  $\hat{\mathcal U}_S $. 

For $h\ne 0$ and if $\Psi$ does not vanish on the $3$-axis then the partial derivatives of $\Psi_N$ and
$\Psi_S=\e^{2\,\ir\,h\,\varphi(p)}\,\Psi_N$ are not both
square integrable. Well defined in $\mathcal M_0 $ and skew hermitian with respect to the measure $\dt p$ are the \idx{covariant derivative}s 
\begin{equation}
D_i  = \ir\, |\vec p|^{-1/2}\, M_{0i}|\vec p|^{-1/2} = \partial_{p^i}+ A_i -\frac{p^i}{2|\vec p|^{2}} \ ,
\end{equation}
with the connection $\vec A$ 
in $\mathcal U_N$ and $\mathcal U_S$
 given by 
\begin{equation}
\label{ablcov}
\vec A_N(p)=
 \frac{\ir\,h}{|\vec p|(|\vec p|+p_z)}
\begin{pmatrix}
-p_y\\
\phantom{-}p_x\\
\phantom{-}0
\end{pmatrix} 
\ ,\ 
\vec A_S(p)=
\frac{-\ir \,h}{|\vec p|(|\vec p|-p_z)}
\begin{pmatrix}
-p_y\\
\phantom{-}p_x\\
\phantom{-}0
\end{pmatrix}\ ,
\end{equation}
and related in the $\mathcal U_N \cap \mathcal U_S$
by the transition function
\begin{equation}
D_{S\,i}=  \e^{2\,\ir\,h\,\varphi(p)} D_{N\,i} \,\e^{-2\,\ir\,h\,\varphi(p)}\ .
\end{equation}
The covariant derivative $D_j$ and the momentum $P^i$  do not constitute Heisenberg pairs\index{Heisenberg pair} as 
the commutator $[D_i,D_j]$ yields the field strength of a momentum space monopole of charge $h$ at $p=0$,
\begin{equation}
\begin{gathered}
\label{monopol}
\ [P^i,P^j] = 0\ ,\ [P^i,D_j] = -\delta^i{}_j\ ,\\ 
\ [D_i, D_j]= F_{ij} = \partial_i A_j - \partial_j A_i 
= \ir\, h\,\varepsilon_{ijk}\frac{P^k}{|\vec P|^3}\ .
\end{gathered}
\end{equation}
The geometry of the massless shell of particles with nonvanishing helicity is noncommutative. 

In terms of the covariant derivative the generators of Lorentz transformations (\ref{masselosnord}, \ref{masselossued}) take the 
rotation equivariant form 
\begin{equation}
\label{masseloskovariant}
-\ir M_{ij} = -(P^i D_{j}- P^j D_{i}) - \ir\,h\,\varepsilon_{ijk}\frac{P^k}{|\vec P|}\ ,\ 
-\ir M_{0i} = - |\vec P|^{1/2} D_{i}|\vec P|^{1/2}\ .
\end{equation}
They satisfy the Lorentz algebra on account of (\ref{monopol}) for arbitrary real~$h$.
However, the covariant derivative~$D$ is a skew hermitian operator only if $2h$ is integer.

The integrand $\mathbb F = \frac{1}{2}\dv p^i\dv p^j F_{ij}$, the first Chern class, is a topological
density: Integrated on each surface $\mathcal S$  which is diffeomorphic to a sphere around the apex $p=0$ of the cone $p^0=|\vec p|$
it yields a value 
\begin{equation}
\frac{1}{4\pi}\int_{\mathcal S}\mathbb F =  \ir\,h
\end{equation}
which depends only on the transition function of the bundle. 
The integral remains constant under smooth, local changes
of the connection $A_i(p)$ of the covariant derivative as the integral of $\mathbb F(A)= \dv (\dv\! p^i\, A_i)$
on coordinate patches changes by boundary terms only and they vanish for local changes.

\section{Angular Momentum}

The massless shell  is foliated in spheres with radius $|\vec p|$, $\mathcal M_0 = \mathbb R_+ \times S^2$.
Hence the Hilbert space $\mathcal H(\mathcal M_0)$ is a tensor product $\mathfrak L^2(\mathbb R_+)\otimes \mathcal H_h(S^2)$ 
and~$\mathfrak L^2(\mathbb R_+)$, the space of wave functions of the energy $E=|\vec p|$, is left pointwise invariant under rotations.
In $\mathcal H_h(S^2)$ the SO$(3)$-representation is induced by the representation 
$R_{D_\delta}=\e^{-\ir \,h \delta}$ of rotations around the $3$-axis. 
To be induced by an irreducible representation does not make the representation of SO$(3)$ irreducible. 
Rather $\mathcal H_h $ decomposes into angular momentum multiplets, each characterized by its 
total angular momentum $j$. Such a multiplet contains a state~$\Lambda$ which is annihilated 
by $J_+ = M_{23}+\ir M_{31}$ and by $M_{12}-j$,
\begin{equation}
(M_{12}-j)\Lambda = 0 ,\ (M_{23}+\ir M_{31})\Lambda = 0\ .
\end{equation}
By (\ref{masselosnord}) these are differential equations for $\Lambda_N$.
They become easily solvable if we consider $\Lambda_N$ as a function of
the complex stereographic coordinates 
\begin{equation}
w = u + \ir v = \frac{p_x + \ir p_y}{|\vec p| + p_z}\ ,\ \bar w = u - \ir v = \frac{p_x - \ir p_y}{|\vec p| + p_z}\ ,
\end{equation}
which map the northern domain to $\mathbb C$. Then the differential equations read 
\begin{equation}
(w \partial_w - \bar w \partial_{\bar w} + h-j) \Lambda_N = 0\ ,\ (w^2\partial_w + \partial_{\bar w} + h\,w)\Lambda_N = 0\ .
\end{equation}

Recollecting that $w\partial_w$ measures the homogeneity in $w$, $w\partial_w\, w^r = r \,w^r$ (where superscripts denote exponents), 
the first equation is solved by $\Lambda_N = w^{j-h}\,g(|w|^2)$ and by the second equation $g$ is
homogeneous in $\bigl(1+|w|^2\bigr)$ of degree $-j$, 
\begin{equation}
\begin{gathered}
\bigl((j - h + h)g + (|w|^2 + 1) g^\prime \bigr)\,w^{j-h+1} = 0\ ,\ 
\label{jstaten}
\Lambda_N(w,\bar w) = \frac{w^{j-h}}{(1+|w|^2)^j}\ .
\end{gathered}
\end{equation}
The state $\Lambda$ is smooth only if $j - h$ is a nonnegative integer. 
It is square integrable  with respect to the rotation invariant measure 
\begin{equation}
\dv \Omega= \frac{4\,\dv u\,\dv v}{(1+u^2+v^2)^2}
\end{equation}
if also $j+h $ is nonnegative which is just the restriction to
be smooth also in southern stereographic coordinates. They are related to the northern coordinates in their 
common domain by inversion at the unit circle, 
\begin{equation}
w^\prime = \frac{p_x + \ir p_y}{|\vec p| - p_z}= \frac{1}{\bar w} \ ,\ 
\bar w^\prime = \frac{p_x - \ir p_y}{|\vec p| - p_z}=\frac{1}{w}\ ,
\end{equation}
and $\Lambda_S$ is smooth only if $j +h$ is a nonnegative integer
\begin{equation}
\label{jstates}
\Lambda_S(w^\prime,\bar w^\prime) = (\frac{w}{| w|}\bigr)^{2h} \Lambda_N =  \frac{w^{\prime\,j+h}}{(1+|w^\prime|^2)^j}\ .
\end{equation}
So $j\ge |h|$: There is no round photon with $j=0$. This follows also from (\ref{helizit}) which 
for $h\ne 0$ excludes that all angular momenta $M_{ij}$ vanish.

The SO$(2)$ representation $R_{D_\delta} = \e^{-\ir h \delta}$, $2h \in \mathbb Z$, induces in the space of sections over the sphere
no SO$(3)$ multiplet with $j<|h|$ and one multiplet for $j = |h|, |h| +1, \dots$, i.e.  the representation $j$ is induced with multiplicity
\begin{equation}
\label{jhel}
n_{h}(j)= 
\left \{
\begin{array}{l l c}
0 & \text{if} & j < |h|\\
1 & \text{if} & j \in \set{|h|, |h|+1, |h|+2,\dots } 
\end{array}
\right .\ . 
\end{equation}
Vice versa, the restriction of the SO$(3)$ representation $j$ contains the SO$(2)$ representation $h$ with the multiplicity
\begin{equation}
m_{j}(h)= 
\left \{
\begin{array}{l l c}
0 & \text{if} & j < |h|\\
1 & \text{if} & j \in \set{|h|, |h|+1, |h|+2,\dots } 
\end{array}
\right .\ . 
\end{equation}
These multiplicities exemplify \index{Frobenius reciprocity} Frobenius' reciprocity \cite{mackey}: 
The representation~$h$ of the subgroup~$H$ induces on sections over $G/H$ each representation~$j$
of the group $G$  with the same multiplicity with which the restriction of $j$ to the subgroup~$H$
contains~$h$, 
\begin{equation}
m_j(h)=n_h(j)\ .
\end{equation}

Photons have helicity $+1$ or $-1$. By Bose symmetry two-photon states to satisfy  
$\Psi^{ij}(p_1,p_2)=\Psi^{ji}(p_2,p_1)$ or
\begin{equation}
\chi^{ij}(u,q) =\chi^{ji}(u,-q)\ ,\ i,j\in\set{+1,-1}\ ,
\end{equation}
where $\chi^{ij}(u,q)$ is the wave function $\Psi^{ij}(p_1,p_2)$ in terms of the center variables $u=(p_1 + p_2)/\sqrt{(p_1 + p_2)^2}$
and the relative momentum \cite{dragon3}
\begin{equation}
q = L_u^{-1}\bigl( p_1 - u (u \cdot p_1)\bigr) = - L_u^{-1}\bigl(p_2 - u (u \cdot p_2)\bigr)\ .
\end{equation}
$L_u^{-1}$ boosts $u$ to $\underline u = (1,0,0,0)$ such that $q\in \mathbb R^3$.
The helicity $i$ is the angular momentum $m_{q}$ of the first photon in the direction of $q$ and $j$ is the angular momentum 
of the second photon in the opposite direction~$-q$ (\ref{helizit}). 
Hence the helicities add to angular momentum $m= i-j$ and
$\chi^{++}$ and $\chi^{--}$ induce rotation multiplets from $m=0$.
As they are two even functions of $q$ they induce two multiplets with even spin $j=0,2,4,\dots$ 

The function $\chi^{+-}$ determines $\chi^{-+}$. Its helicities 
combine to angular momentum  $m=2$ which by (\ref{jhel}) induces one
multiplet with spin $|j| = 2, 3,\dots$. 

So there is no spin-1 multiplet of two photons. 
This conclusion is the Landau Yang theorem \cite{landau,yang}.
It clarifies why positronium
with $j=1$ does not decay into two photons: 
the interaction is Poincaré invariant and therefore preserves the total momentum $P$ and the total spin. 
But there is no two-photon state with 
$j=1$ into which $j=1$ positronium can decay. 
There is a three photon state with $j=1$. But the decay rate to three photons  is suppressed roughly
by a factor $\alpha\sim 1/137$ for the production of an additional photon which is why one finds positronium predominantly in its stablest
state as orthopositronium with $j=1$.

\section{Position of a Massless Particle}

For a massless particle there cannot exist a spatial position operator $\vec X$ which shifts the momentum,
$\e^{\ir\,\vec b\,\vec X}\vec P\, \e^{-\ir\,\vec b\,\vec X}= \vec P + \vec b$,  and generates together with \mbox{$P^0=\sqrt{\vec P^2}$} an 
algebra. Such an algebra would contain  
\begin{equation}
[X^i,[X^i, P^0]]= - \frac{D-2}{|\vec P|}\ , 
\end{equation}
thus, for $D\ne 2$, 
all inverse powers of $|\vec P|$. Hence the domain of the algebra contained only states $\Psi$ 
which decreased near $\vec p=0$ faster than any power of $|\vec p|$. As the domain of the generators  is invariant  under the group,
it is invariant under shifts and contained all shifted states. So 
$(\e^{\ir\,\vec b\,\vec X}\Psi)(\vec p) = \Psi(\vec p + \vec b)$ would have to vanish for all $\vec b$ at $\vec p=0$. But this means 
$\Psi=0$. The Heisenberg algebra of $\vec X$ and $\vec P$ is incompatible with the Hamiltonian $\sqrt{\vec P^2}$, which is not a smooth operator but rough at $\vec p=0$.

In the plane of spatial momenta of the massless shell the Lorentz fixpoint \mbox{$\vec p = 0$} breaks the translation symmetry of the plane.
There the dispersion  $p^0=\sqrt{\vec p^2}$ is not smooth. 
This lack of translational symmetry does not occur on massive shells and frustrates all attempts \cite{hawton,newton, pryce, wightman}
to construct a position operator for massless particles.

\section{Conclusions}
We corrected widespread misconceptions about the hermitian generators of Poincaré transformations. Using the traditional notation 
of quantum physics we reconciled the smoothness requirements of physical states with the singularities which the Lorentz generators,
acting on massless states of nonvanishing helicity, 
develop on the negative $z$-axis (common representations are even singular on the complete $z$-axis). The singularity vanishes on the negative $z$-axis
in a different gauge but reappears on the positive $z$-axis: the smooth wave function is a section of a bundle which does not allow globally 
partial derivatives with respect to the momenta, but only noncommuting covariant derivatives. This constitutes arguably the most elementary example of 
noncommutative geo\-me\-try
in physics.

The modulus of the helicity is a lower bound of the total angular momentum of one-photon states. It excludes a two-photon states with $J=1$ confirming 
the Landau Yang theorem, which prevents orthopositronium to decay into two photons. 

At the fixpoint $p=0$ of Lorentz transformations the Hamiltonian $H=\sqrt{\vec p^2}$ is only continuous, not smooth. This is why one cannot extend the algebra of 
Poincaré generators by a spatial position operator, which generates translations of spatial momentum.

\begin{appendices}

\section{The Wigner Rotation of Massive Particles}\label{secA1}

To derive the angle of the Wigner rotations, we use the fact that the Lorentz group SO$(1,3)$ represents SL$2,\mathbb C)$.
For definiteness and to fix our notation we review these well-known relations.
We use the $\eins$-matrix $\sigma^0$ and the  Pauli matrices $\sigma^i$, $i=1,2,3$, 
with matrix elements $\sigma^m_{\alpha\dot \beta}$, $m\in \set{0,1,2,3}$, $\alpha,\dot \beta \in \set{1,2}$,
\begin{equation}
\index{$\sigma^m, \bar \sigma^n$}\index{Pauli matrices}
\label{pauli}
\sigma^0 =
\begin{pmatrix}
1&\\
&1
\end{pmatrix}\ ,\ 
\sigma^1 =
\begin{pmatrix}
&1\\
1&
\end{pmatrix}\ ,\ 
\sigma^2 =
\begin{pmatrix}
&-\ir \\
\ir &
\end{pmatrix}\ ,\ 
\sigma^3 =
\begin{pmatrix}
1&\\
&-1
\end{pmatrix}
\end{equation}
and write products 
\begin{equation}
\label{paulialgebra}
\sigma^i \sigma^j = \delta^{ij}{\eins }+ \ir\, \varepsilon_{ijk}\sigma^k\ ,\quad i,j,k\in\{1,2,3\}\ ,\ \varepsilon_{123}= 1
\end{equation}
of their linear combinations concisely as
\begin{equation}
\label{paulialgvec}
(\vec{m}\cdot \vec{\sigma})( \vec{n}\cdot \vec{\sigma}) = 
(\vec{m}\cdot\vec{n}){\eins }+ \ir\, (\vec{m}\times \vec{n})\cdot \vec{\sigma}\ .
\end{equation}
If $\vec m \parallel \vec n$ then $\vec{m}\cdot \vec{\sigma}$ and $\vec{n}\cdot \vec{\sigma}$ commute, 
they anticommute if $\vec{m} \perp \vec{n}$.

For each unit vector $\vec{n}$ one has $(\vec{n}\cdot \vec{\sigma})^2 = {\eins }$
and series of $\vec{n}\cdot \vec{\sigma}$ simplify
\begin{align}
\label{uvonalpha}
U_{\alpha \vec n}=\exp(-\ir\frac{\alpha}{2}\, \vec{n}\cdot \vec{\sigma})&=
\bigl(\cos \frac{\alpha}{2}\bigr)\,\eins  -\ir \,\bigl( \sin\frac{\alpha}{2}\bigr)\, \vec{n}\cdot \vec{\sigma}\ ,\\
\label{eh}
V_{\beta\vec n}=\exp (-\frac{\beta}{2}\vec{n}\cdot \vec{\sigma})&= (\ch\frac{\beta}{2})\eins-
(\sh\frac{\beta}{2})\, \vec{n}\cdot \vec{\sigma}\ .
\end{align}

The matrices $\sigma^m$ are a basis of the real four dimensional vectorspace of hermitian
2$\times$2-matrices 
\begin{equation}
\hat k= \sigma^m k^n \eta_{mn} = 
\begin{pmatrix}
k^0 - k^3& -k^1 + \ir k^2\\
-k^1 - \ir k^2& k^0 + k^3
\end{pmatrix}\ .
\end{equation}
Their determinant $\det \hat k = k^m k^n \eta_{mn}= (k^0)^2 - (k^1)^2 - (k^2)^2 - (k^3)^2$
is invariant for all $D$ with $\det D = 1$ under the linear transformation ($D^\dagger \coloneqq D^{*\T}$)
\begin{equation}
\hat{k}\mapsto \hat{k'} = D\, \hat k\,D^\dagger\ ,\ k'=\Lambda_D k\ .
\end{equation}
So $k\mapsto k'= \Lambda_D k$ is a Lorentz transformation and 
$D \mapsto \Lambda_D$ is a representation of SL$(2,\mathbb C)$ in SO$(1,3)$, 
$\Lambda_D\Lambda_{D'}=\Lambda_{DD'}$. Physics parlance reverses the relation and claims sloppily that $D$ represents~$\Lambda$.

Explicitly $U_{\alpha\vec n}^\dagger=U_{\alpha\vec n}^{-1}$ is unitary and effects a rotation by $\alpha$ around $\vec n$.
To see this decompose $\vec{k}=(\vec k\cdot \vec n)\,\vec{n}+\vec{k}_\perp$ into parallel and transverse parts
and split $k = k_\parallel + k_\perp$ with $k_\parallel=(k^0, (\vec k \cdot \vec n)\vec n)$
and $k_\perp=(0,\vec k_\perp)$.

$\hat k_\parallel$ commutes with each power series in $\vec{n}\cdot \vec{\sigma}$.
Hence it is invariant 
\begin{equation}
U \,\hat k_\parallel\, U^\dagger= \hat k_\parallel\, U\, U^\dagger = \hat k_\parallel\ .
\end{equation}
As $\vec k_\perp$ is orthogonal to $\vec n$ the matrix
$\hat k_\perp$ anticommutes with  $\vec{n}\cdot \vec{\sigma}$ (\ref{paulialgvec})
\begin{equation}
\label{antivert}
\hat k_\perp\  \vec{n}\cdot \vec{\sigma} = - \vec{n}\cdot \vec{\sigma}\ \hat k_\perp\ .
\end{equation}
Therefore $U (\vec{k}_\perp  \cdot \vec{\sigma}) U^\dagger =U^2 (\vec{k}_\perp  \cdot \vec{\sigma}) $
and $U_{\alpha\vec n}^2=U_{2\alpha\,\vec n}$ implies
\begin{equation}
U (\vec{k}_\perp  \cdot \vec{\sigma}) U^\dagger 
= (\cos\alpha -\ir \sin\alpha\,\vec{n}\cdot \vec{\sigma})(\vec{k}_\perp\cdot \vec{\sigma})
 \stackrel{(\ref{paulialgvec})}{=} (\cos \alpha\, \vec{k}_\perp+ \sin\alpha\, \vec{n}\times \vec{k}_\perp )\cdot  \vec{\sigma} .
\end{equation}
So $U_{\alpha\vec n}=\e^{-\ir \frac{\alpha}{2}\vec{n} \cdot \vec{\sigma}}$ causes by
$\hat k\mapsto U_{\alpha\vec n}\,\hat k\,U_{\alpha\vec n}^{\dagger}$
the rotation $k \mapsto D_{\alpha \vec n}k$ of four-vectors~$k$ by the angle $\alpha$ around the axis $\vec{n}$ 
\begin{equation}
\label{dvonu}
D_{\alpha\vec n}:\ {k}_\parallel+k_\perp\mapsto 
k_\parallel+ \cos\alpha\, k_\perp+ \sin\alpha\,(0,\vec{n}\times\vec{k}_\perp) .
\end{equation}
Vice versa, to the rotation $D_{\alpha\,\vec{n}}$ by~$\alpha$  around the axis $\vec{n}$  there corresponds
the pair of unitary matrices $U=\e^{-\ir \frac{\alpha}{2}\vec{n}\cdot  \vec{\sigma}}$ 
and $-U=\e^{-\ir \frac{\alpha+2\pi}{2}\vec{n}\cdot  \vec{\sigma}}$. 

\longpage

As $V_{\beta\vec n}=V_{\beta\vec n}^\dagger$ is hermitian and because $\hat k_\perp$
anticommutes with  $\vec{n}\cdot \vec{\sigma}$, the transverse part is invariant
\begin{equation}
V (\vec{k}_\perp  \cdot \vec{\sigma}) V^\dagger = V V^{-1} (\vec{k}_\perp  \cdot \vec{\sigma}) =\vec{k}_\perp  \cdot \vec{\sigma}
\end{equation}
while the longitudinal part $\hat k_\parallel=k^0 - (\vec k \cdot \vec n) \vec n\cdot  \vec \sigma$ commutes and 
by $V^2_{\beta\vec n}\stackrel{\ref{eh}}{=} V_{2\beta \vec n}$ transforms as 
\begin{align}
\nonumber
V_{\beta\vec n} \hat{k}_\parallel\, V^\dagger_{\beta\vec n} &= \hat{k}_\parallel\, V_{2\beta\vec n}
=(k^0\eins - (\vec k \cdot \vec n)\, \vec n\cdot \vec{\sigma} ) 
 (\ch\beta \eins - (\sh\beta) \vec{n}\cdot \vec{\sigma})\\
\nonumber
& \stackrel{\ref{paulialgvec}}{=}
\bigl ((\ch\beta)\, k^0 + (\sh\beta)\, (\vec k \cdot \vec n)\bigr ) \eins
- \bigl ( (\sh\beta)\, k^0 + (\ch\beta)\, (\vec k \cdot \vec n)\bigr )\vec{n}\cdot \vec{\sigma}\\
& = k^{\prime\, 0}\eins - (\vec k^\prime\cdot \vec n)\,\vec n \cdot \vec{\sigma}\ .
\end{align}
We conclude 
\begin{equation}
\vec k^{\prime}_{\perp}=\vec k_\perp \ ,\ 
\begin{pmatrix}
k^{\prime\,0}\\
(\vec k^{\prime}\cdot \vec n)\,\vec n
\end{pmatrix}=
\begin{pmatrix}
\ch\beta & \sh\beta \, \vec n^{\T}\\
\sh\beta\,\vec n & \ch\beta\, \vec n\, \vec n^{\T}
\end{pmatrix}
\begin{pmatrix}
k^{0}\\
(\vec k\cdot \vec n)\,\vec n
\end{pmatrix}\ .
\end{equation}
So $V_{\beta \vec n} $ effects the \idx{boost} $L_u$ (\ref{lorp}) 
in direction of~$\vec n = \vec u / |\vec u |$ with rapidity~$\beta$, $\sh \beta= |\vec u|$. 
To each boost $L_u$ there corresponds the pair of $V_{\beta \vec n} $ and 
$-V_{\beta \vec n} $ of SL$(2,\mathbb C)$ matrices.
SL$(2,\mathbb C)$ is a double cover of SO$(1,3)^\uparrow$.

To calculate the Wigner rotation for massive particles 
\begin{equation}
W({\Lambda,p}) = L_{\Lambda u}^{-1} \Lambda L_u \ ,\ p = m u
\end{equation}
is a \lq Herculean task\rq\  \cite{ungar1} and requires \lq tedious manipulations\rq\  \cite{sexl}.
As we are unaware of a citable derivation, we present here a derivation. 
It simply reads the Wigner rotation from the product of SL$(2,\mathbb C)$ matrices $\lambda$, $l_u$ and~$w$
\begin{equation}
\lambda l_u  = l_{\Lambda u} w(\Lambda,p)
\end{equation}
which correspond to the Lorentz transformation $\Lambda$, the boosts $L_u$, $L_{\Lambda u}$
and the Wigner rotation. Using the notation
\begin{equation}
l_a = \ch a^\prime - \sh a^\prime\, \vec n_a\cdot \sigma \ ,\ 
w = \cos \delta^\prime - \ir\,\sin \delta^\prime\, \vec n \cdot \sigma  \ ,\ 
a^\prime = \frac{a}{2}  \ ,\ \delta^\prime = \frac{\delta}{2}\ ,
\end{equation}
the products of two boosts $l_b l_a$  and of a boost $l_{-c}$ with a rotation~$w$ yield
\begin{equation}
\begin{gathered}
\bigl(\ch b^\prime - \sh b^\prime\,\vec n_b\cdot \sigma \bigr)
\bigl(\ch a^\prime - \sh a^\prime\,\vec n_a\cdot  \sigma \bigr)=\\
(\ch b^\prime \ch a^\prime + \sh b^\prime \sh a^\prime\, \vec n_a\cdot \vec n_b)\eins  - 
(\ch a^\prime\, \sh b^\prime \,\vec n_b + \ch b^\prime\, \sh a^\prime\, \vec n_a)\cdot \sigma\, + \\
+ \ir \,\sh b^\prime\, \sh a^\prime\,  (\vec n_b\times  \vec n_a)\cdot \sigma \ ,
\end{gathered}
\end{equation}
\begin{equation}
\begin{gathered}
\bigl(\ch c^\prime + \vec n_c\cdot \sigma \sh c^\prime\bigr)
\bigl(\cos \delta^\prime - \ir \sin \delta^\prime\, \vec n\cdot \sigma \bigr)=\\
(\ch c^\prime \cos\delta^\prime - \ir \sh c^\prime\, \sin \delta^\prime \,\vec n_c\cdot \vec n)\eins
+ \sh c^\prime ( \cos \delta^\prime \vec n_c +  \sin \delta^\prime\, \vec n_c \times \vec n )\cdot \sigma\, + \\
-\ir \,\ch c^\prime \sin \delta^\prime \,\vec n\cdot \sigma\ .
\end{gathered}
\end{equation}

Both products are equal, $l_b l_a = l_{-c} w$, if the complex coefficients of the linearly independent matrices $\eins$ and the
$\sigma$-matrices  match. Comparing the coefficients determines~$w$ without knowing $l_{-c}$ explicitly.

The coefficients of~$\eins$ agree only if  $\vec n\cdot \vec n_c=0$. The axis $\vec n$ of the Wigner rotation 
is orthogonal to the direction $\vec n_c$ of the resulting boost; $\vec n_c$ lies in the plane spanned by
$\vec n_a$ and $\vec n_b$, as it must: $l_b$ and~$l_a$ 
are in the subgroup SO$(1,2)$ of boosts and rotations in this plane.

With $\vec n_b \cdot \vec n_a = \cos \gamma$ and $\vec n_a \times \vec n_b = \vec n\,\sin \gamma$, where 
$\gamma$ is the angle included by $\vec n_a$ and $\vec n_b$,
the comparison of the coefficients of~$\eins$ and of $\vec n\cdot \sigma$ yields
\begin{equation}
\begin{aligned}
(\ch c^\prime)\,(\cos \delta^\prime)&= (\ch b^\prime)(\ch a^\prime) + (\sh b^\prime)(\sh a^\prime)(\cos \gamma)\ ,\\ 
(\ch c^\prime)\,(\sin \delta^\prime)&= (\sh b^\prime)(\sh a^\prime)(\sin \gamma)\ .
\end{aligned}
\end{equation}
The ratio of both equations determines the looked for angle $\delta=2\delta^\prime$ \cite{ungar1}
\begin{equation}
\label{wignerdelta}
\tan \frac{\delta}{2}= \frac{\sin \gamma}{\bigl(\coth \frac{a}{2}\,\coth \frac{b}{2}\bigr)+\cos\gamma} 
\ .
\end{equation}
It has the same sign as $\gamma$, the Wigner rotation $W({L_u,p})$ rotates from $\vec p$ to~$\vec u$.

In the hyperbolic plane
\footnote{In $H^d$, $d > 2$, the deficit angles of several nonplanar triangles  cannot simply be added, because the corresponding 
rotations around different axes do not commute.
}
the area of the hyperbolic triangle is this deficit angle $\delta=\pi - \alpha -\beta -\gamma$ \cite{gauss}. 
The area is bounded by~$\pi$ though the sides of the hyperbolic triangle can be arbitrarily long. 

\section{Iwasawa Decomposition}
By the Schmidt orthogonalization procedure 
each matrix $g\in\text{SL}(d,\mathbb C)$ decomposes continuously into a product of 3 factors, $g=kan$, where
$k$ is from $K=\text{SU}(d)$, which is compact,  $a$ is real and diagonal and from a noncompact abelian subgroup~$A$ 
and $n$ is lower triangular from a subgroup $N$ of exponentials of nilpotent matrices. Elementary algebra 
confirms such an \idx{Iwasawa decomposition} of $g\in\text{SL}(2,\mathbb C)$ to be 
\begin{equation}
\label{kan}
\begin{pmatrix}
a & c\\
b & d
\end{pmatrix}
=
\begin{pmatrix}
\phantom{-}\delta^* & \beta \\
-\beta^* & \delta
\end{pmatrix}
\begin{pmatrix}
1/ r & 0\\
0  & r
\end{pmatrix}
\begin{pmatrix}
1 & 0\\
z & 1
\end{pmatrix}
\end{equation}
where $ad-bc = 1$ and 
\begin{equation}
\label{e2transz}
r = \sqrt{|c|^2 + |d|^2}\ ,\ \beta = \frac{c}{r}\ ,\ \delta = \frac{d}{r}\ ,\  z = \frac{a c^* + b d^*}{r^2}\ . 
\end{equation}
The first factor in the decomposition corresponds to a rotation, the second to a boost in $3$-direction and the third matrix 
represents an E$(2)$-translation (\ref{hgruppe}).

Decompositions of the group SL$(d,\mathbb C)=KAN$ into 
into a product of a maximal compact subgroup~$K$, an abelian subgroup $A$ and a nilpotent subgroup $N$ 
are equivalent, $K'=sKs^{-1}$, $A'=sKs^{-1}$ and $N'=sNs^{-1}$.
For fixed groups $K$, $A$ and $N$ each group element $g=kan$ factorizes uniquely.

\end{appendices}



\begin{thebibliography}{99.}%
\bibitem{bose}Samir K. Bose and R. Parker, \emph{Zero-Mass Representation of Poincaré Group and Conformal Invariance,} J. Math. Phys. 10 (1969) 812--813
\bibitem{dimock}Jonathan Dimock, \emph{Locality in Free String Field Theory-II,} Annales Henri Poincaré 3 (2002) 613 
\url{http://arxiv.org/abs/math-ph/0102027}
\bibitem{dixmier}Jacques Dixmier and Paul Malliavin, \emph{Factorisations de fonctions et de vecteurs indéfiniment différentiables,}
 Bull. Sci. Math. 102 (1978) 305--330
\bibitem{dragon1}Norbert Dragon,\emph{ The Geometry of Special Relativity -- a Concise Course,} Springer, Berlin, 2012
\bibitem{dragon3}Norbert Dragon, \emph{Relativistic Covariance of Scattering,} \url{https://arxiv.org/abs/2307.15426}
\bibitem{dragono1}Norbert Dragon and Florian Oppermann, \emph{Heisenberg versus the Covariant String,} \url{https://arxiv.org/abs/2212.07256}
\bibitem{dragono2}Norbert Dragon and Florian Oppermann, \emph{The Rough with the Smooth of the Light Cone String,} \url{https://arxiv.org/abs/2212.14822}
\bibitem{ek}Bengt Ek and Bengt Nagel, \emph{Differentiable vectors and sharp momentum states of helicity representations of the Poincaré group,}
J. Math. Phys. 25 (1984) 1662--1670 
\bibitem{fronsdal}Moshé Flato, Christian Fronsdal and Daniel Sternheimer, \emph{Difficulties with massless particles?,} Comm. Math. Phys. 90 (1983) 563--573 
\bibitem{gauss}\emph{Carl Friederich Gauss Werke, Achter Band,} Springer, Berlin (1900) 220--224, Brief an Wolfgang von Bolyai, G\"ottingen,  6. M\"arz 1833
\bibitem{hatcher}Allen Hatcher, \emph{Algebraic Topology,} Cambridge University Press, 2002
\bibitem{hawton}Margret Hawton, \emph{Position Operator with Commuting Components,} Phys. Rev. A 59 (1999) 954--959
\bibitem{landau}Lew Landau, \emph{On the angular momentum of a system of two photons,}  Doklady Akademii Nauk Ser. Fiz. 60 (1948) 207--209
\bibitem{lomont}John S. Lomont and Harry E. Moses, 
\emph{Simple Realizations of the Infinitesimal Generators of the Proper Orthochronous Imhomogeneous Lorentz Group for Mass Zero,}
J. Math. Phys. 3 (1962) 405--408
\bibitem{mackey}George W. Mackey, \emph{Induced Representations of Locally Compact Groups I,}
Annals of Ma\-the\-matics 55 (1952) 101--139;
\emph{Induced Representations of Locally Compact Groups II,}
Annals of Mathematics 58 (1953) 193--221;  
\emph{Induced Representations of Groups and Quantum Mechanics,} W. A. Benjamin, New York, 1968
\bibitem{mund}Jens Mund, \emph{String-Localized Quantum Fields, Modular Localization, and Gauge Theories,}  
Talk given at the Fifth International Conference on Mathematical Methods in Physics IC2006, PoS(IC2006)028 
\bibitem{newton}Theodore Duddell Newton and Eugene Paul Wigner, \emph{Localized States for Elementary Systems,}  Rev. Mod. Phys. 21 (1949) 400--406
\bibitem{pryce}Maurice H. L. Pryce, \emph{The mass-centre in the restricted theory of re\-la\-ti\-vi\-ty 
and its connexion with the quantum theory of elementary particles,} 
Proc. R. Soc. London Ser. A 195 (1948) 62
\bibitem{reeh}Helmut Reeh and Siegfried Schlieder, \emph{Bemerkungen zur Unitär\-äqui\-valenz von Lorentz\-in\-varianten Feldern,} Il Nuovo Cimento 22 (1961) 1051--1068
\bibitem{richter}David A. Richter,\\ \url{https://commons.wikimedia.org/wiki/File:Hopfkeyrings.jpg}
\bibitem{schmuedgen}Konrad Schm\"udgen, \emph{Unbounded Operator Algebras and Representation Theory,} Birkh\"auser, Basel, 1990,  Chapter 10\\
\emph{An Invitation to Unbounded Representations of $\star$-Algebras on Hilbert Space,} Springer Nature Switzerland, Cham, 2020 
\bibitem{sexl}Roman U. Sexl and Helmuth K. Urbantke, \emph{Relativity, Groups, Particles,} Springer,  Wien, 2001
\bibitem{ungar1}Abraham A. Ungar, \emph{Beyond the Einstein Addition Law and its Gyroscopic Thomas Precession,}
 Kluwer Academic Publishers, Dordrecht, 2001
\bibitem{weinberg}Steven Weinberg, \emph{The Quantum Theory of Fields,} Cambridge University Press, 1995
\bibitem{wightman}Arthur S. Wightman, \emph{On the Localizability of Quantum Mechanical Systems,} Rev. Mod. Phys. 34 (1962) 845--872
\bibitem{wigner1948}Eugene Paul Wigner, \emph{Relativistische Wellengleichungen,} Zeitschrift für Physik A, 124 (1948)  665--684\\
\emph{Gruppentheorie und ihre Anwendung auf die Quantenmechanik der Atomspektren,} Vieweg Verlag, Braunschweig, 1931
\bibitem{yang}Chen Ning Yang, \emph{Selection Rules for the Dematerialization of a Particle into Two Photons,}
Phys. Rev. 77 (1950) 242--245
\end{thebibliography}
\end{document}